\documentclass{article}
\usepackage{amsfonts}

\usepackage{amssymb}
\usepackage{amsmath}
\usepackage{cite}

\begin{document}

\title{Symmetries of Differential Equations in Cosmology}
\author{Michael Tsamparlis$^{1}$ and Andronikos Paliathanasis$^{2,3,4}$ \\
\textit{\ }$^{1}$\textit{\small Department of Physics, Section of
Astrophysics-Astronomy-Mechanics, } \\
\textit{\small University of Athens, Zografos 15783, Athens, Greece}\\
$^{2}$\textit{{\small \textit{Instituto de Ciencias F\'{\i}sicas y Matem\'{a}%
ticas, Universidad Austral de Chile, Valdivia, Chile}}}\\
$^{3}$\textit{{\small \textit{Department of Mathematics and Natural
Sciences, Core Curriculum Program,}}}\\
\textit{{\small \textit{\ Prince Mohammad Bin Fahd University, Al Khobar
31952, Kingdom of Saudi Arabia}}}\\
$^{4}$\textit{{\small \textit{Institute of Systems Science, Durban
University of Technology, }}}\\
\textit{{\small \textit{PO Box 1334, Durban 4000, Republic of South Africa }}%
}}
\date{}
\maketitle

\begin{abstract}
The purpose of the current article is to present a brief albeit accurate
presentation of the main tools used in the study of symmetries of Lagrange
equations for holonomic systems and subsequently to show how these tools are
applied in the major models of modern cosmology in order to derive exact
solutions and deal with the problem of dark matter/energy. The key role in
this approach are the first integrals of the field equations. We start with
the Lie point symmetries and the first integrals defined by them, that is
the Hojman integrals. Subsequently we discuss the Noether point symmetries
and the well known method for deriving the Noether integrals. By means of
the Inverse Noether Theorem we show that to every Hojman quadratic first
integral one is possible to associate a Noether symmetry whose Noether
integral is the original Hojman integral. It is emphasized that the point
transformation generating this Noether symmetry need not coincide with the
point transformation defining the Lie symmetry which produces the Hojman
integral. We discuss the close connection between the Lie point and the
Noether point symmetries with the collineations of the metric defined by the
kinetic energy of the Lagrangian. In particular the generators of Noether
point symmetries are elements of the homothetic algebra of that metric. The
key point in the current study of cosmological models is the introduction of
the mini superspace, that is the space which is defined by the physical
variables of the model, which is not the spacetime where the model evolves.
The metric in the mini superspace is found from the kinematic part of the
Lagrangian and we call it the kinetic metric. The rest part of the
Lagrangian is the effective potential. We consider coordinate
transformations of the original mini superspace metric in order to bring it
to a form where we know its collineations that is, the Killing vectors, the
homothetic vector etc. Then we write the field equations of the cosmological
model and we use the connection of these equations with the collineations of
the mini superspace metric to compute the first integrals and subsequently
to obtain analytic solutions for various allowable potentials and finally
draw conclusions about the problem of dark energy. We consider the $\Lambda
CDM$ cosmological model, the scalar field cosmology, the Brans Dicke
cosmology, the $f(R)$ gravity, the two scalar fields cosmology with
interacting scalar fields and the Galilean cosmology. In each case we
present the relevant results in the form of Tables for easy reference.
Finally we discuss briefly the higher order symmetries (the contact
symmetries) and show how they are applied in the cases of scalar field
cosmology and in the $f(R)$ gravity \newline
\end{abstract}

\section{Introduction}

\label{intro}

In order to understand properly the role of symmetries in Cosmology we have
to make a short detour into General Relativity and the relativistic models
in general. General Relativity associates the gravitational field with the
geometry of spacetime as this is specified by the metric of the Riemannian
structure. Concerning the matter this is described in terms of various
dynamical fields which are related to the mater via Einstein equations $%
G_{ab}=T_{ab}$ (we consider that the Einstein's gravitational constant is $k$%
=1). Einstein equations are not equations, in the sense that they equate
known quantities in terms of unknown ones, that is there is no point to look
for a "solution" of them in this form. These equations are rather generators
of equations which result after one introduces certain assumptions according
to the model required. These assumptions are of two kinds: a) Geometric
assumptions; b) Other non-geometric assumptions among the physical fields
which we call equations of state. The first specify the metric to a certain
degree and are called collineations \cite{sym5} (or "symmetries" \ which is
in common use and is possible to create confusion in our discussion of
symmetries in Cosmology). The collineations are the familiar Killing vectors
(KV), the conformal Killing vectors (CKV), etc. The collineations affect the
Einstein tensor which is expressed in terms of the metric. For example for a
KV $X$ one has $L_{X}g_{ab}=L_{X}G_{ab}=0$.

Through Einstein field equations collineations pass over to $T_{ab}$ and
restrict its the possible forms, therefore the types of matter it can be
described in the given geometry-model. For example if one considers the
symmetries of the Friedmann Robertson Walker (FRW) model, which we shall
discuss in the following, then the collineations restrict the $T_{ab}$ to be
of the form
\begin{equation}
T_{ab}=\rho u_{a}u_{b}+ph_{ab}
\end{equation}%
where $\rho $,$p$ are two dynamical scalar fields the density \ and the
isotropic pressure, $u^{a}$ are the comoving observers $u^{a}$ ($%
u^{a}u_{a}=-1)$ and $h_{ab}=g_{ab}+u_{a}u_{b}$ is the spatial projection
operator. An equation of state is a relation between the dynamical variables
$\rho ,p.$

Once one specifies the metric by the considered collineations (and perhaps
some additional requirements of geometric nature) of the model and
consequently the dynamical fields in the energy momentum tensor then
Einstein equations provide a set of differential equations which describe
the defined relativistic model. What it remains is the solution of these
equations and the consequent determination of the Physics of the model. At
this point one introduces the equations of state which simplify further the
resulting field equations.

When one has the final form of the field equations enters a second use of
the concept of \textquotedblright symmetry\textquotedblright\ which is the
main objective of the current work. Let us refer briefly some history.

In the late of 19th century Sophus Lie in a series of works \cite%
{lie1,lie2,lie3} with the title \textquotedblleft Theory of transformation
groups\textquotedblright\ introduced a new method for the solution for
differential equations via the concept of \textquotedblright
symmetry\textquotedblright . In particular Lie defined the concept of
symmetry of a differential equation by the requirement the point
transformation leaves invariant the set of solution curves of the equation,
that is, under the action of the transformation a point form one solution
curve is mapped to a point in another solution curve. Subsequently Lie
introduced a simple algebraic algorithm for the determination of this type
of symmetries and consequently on the solution of differential equations.
Since then, symmetries of differential equations is one of the main methods
which is used for the determination of solutions for differential equations.
Some important works which established the importance of symmetries in the
scientific society are those of Ovsiannikov \cite{Ovsi}, Bluman and Kumei
\cite{kumei}, Ibragimov \cite{ibra}, Olver \cite{olver}, Crampin \cite%
{Crampin}, Kalotas \cite{kalotas} and many others; for instance see \cite%
{prince,luz,leach1,leach2,mahomed,nucci,harrison,book1,popo,katz2}.

As it is well known an important fact in the solution of a differential
equation are the first integrals. Inspired by the work of Lie in the early
years of the 20th century, Emmy Noether required another definition of
symmetry, which is known as Noether symmetry, which concerns Lagrangian
dynamical systems and it is defined by the requirement that the action
integral under the action of the point transformation changes up to a total
derivative so that Lagrange equation(s) remain the same\cite{noe1}. She
established a connection between the Noether symmetries and the existence of
a first integral which she expressed by a simple mathematical formula. In
addition, Noether's work except of its simplicity had a second novelty by
alloying the continuous transformation to depend also on the derivatives of
the dependent function, which was the first generalization of the context of
symmetry from point transformations to higher-order transformations.
Symmetries play an important role in various theories of physics, from
analytical mechanics \cite{sym1}, to particle physics \cite{sym2,sym4}, and
gravitational physics \cite{sym3,sym5}.

In the following we shall present briefly the approach of Lie and Noether
symmetries and will show how the symmetries of differential equations are
related to the collineations of the metric in superspace. We shall develop
an algorithm which indicates how one should work in order to get the
analytical solution of a cosmological model. The various examples will
demonstrate the application of this algorithm.

In conclusion, by symmetries in Cosmology we mean the work of Lie and
Noether applied to the solution of the field equations of a given
cosmological model - scenario.

\section{Point transformations}

On a manifold $M$ with coordinates $(t,q^{a})$ one defines the jet space $%
J^{m}(M)$ of order $m$ over $M$ to be a manifold with coordinates $t,q^{a},%
\frac{dq^{a}}{dt},...,\frac{d^{m}q^{a}}{dt^{m}}.$ Let $X=\xi \left(
t,q,...,q^{[m]}\right) \frac{\partial }{\partial t}+\eta ^{a[A]}\left(
t,q,...,q^{[m]}\right) \frac{\partial ^{\lbrack A]}}{\partial q^{a[A]}}$
where $A=1,...,m$ be a vector field on $J^{m}(M)$ which generates the
infinitesimal point transformation%
\begin{eqnarray}
t^{\prime } &=&t+\varepsilon \xi +O^{2}\left( \varepsilon ^{2}\right) +\cdots
\notag \\
q^{a^{\prime }} &=&q^{a}+\varepsilon \eta ^{a[1]}+O^{2}\left( \varepsilon
^{2}\right) +\cdots  \notag \\
&&...  \label{PointTrans.1} \\
q^{a^{\prime }[m]} &=&q^{a[m]}+\varepsilon \eta ^{a[m]}+O\left( \varepsilon
^{2}\right) +\cdots  \notag
\end{eqnarray}%
\qquad \qquad

Well behaved infinitesimal point transformations form a group under the
operation of composition of transformations. If the infinitesimal point
transformation depends on many parameters, that is,

\begin{equation}
q^{\prime i}=q^{\prime i}\left( q^{i},q^{i(m)},\mathbf{E}\right) ~~
\label{LA.00}
\end{equation}%
where $\mathbf{E}=\varepsilon ^{\beta }\partial _{\beta },$~is a vector
field in $%
\mathbb{R}
^{\kappa },~\beta =1...\kappa $ with the same properties as in the case of
the one parameter infinitesimal point transformations, then the point
transformation is called a multi-parameter point transformation. These
transformations are generated by $\kappa -$vector fields which form a
(finite or infinite dimensional)\ Lie algebra. That is, if the vector fields
$X,Y$ are generators of a multi-parameter point transformation so is the
commutator $Z~=\left[ X,Y\right] .$

\subsection{Prolongation of point transformations}

A\ differential equation $H\left( t,q,q^{\prime },...,q^{\left( m\right)
}\right) =0~$where $q^{i}(t)$ and $q^{i\left( m\right) }=\frac{d^{m}q^{i}}{%
dt^{m}}$ may be considered as a function on the jet space $J^{m}(M).$ In
order to study the effect of a point transformation in the base manifold $%
M(t,q)$ to the differential equation one has to prolong the transformation
to the space $J^{m}(M).$ To do that we consider in $J^{m}(M)$ the induced
point transformation%
\begin{align*}
\bar{t}& =t+\varepsilon \xi ~,~\bar{y}=x+\varepsilon \eta \\
\bar{q}^{i\left( 1\right) }& =q^{i\left( 1\right) }+\varepsilon X^{\left[ 1%
\right] }~,~.... \\
& ... \\
\bar{q}^{i\left( n\right) }& =q^{i\left( n\right) }+\varepsilon X^{\left[ n%
\right] }
\end{align*}%
where $X^{\left[ k\right] }$ $k=1,2,...,m$ are the components of a vector
field
\begin{equation}
W=\xi \partial _{x}+\eta \partial _{y}+X^{\left[ 1\right] }\partial
_{y^{\left( 1\right) }}+...+X^{\left[ m\right] }\partial _{y^{\left[ m\right]
}}.
\end{equation}%
in $J^{m}(M)$ called the lift of the vector field $X=\xi (t,q)\frac{\partial
}{\partial x^{i}}+\eta ^{i}(t,q)\frac{\partial }{\partial q^{i}}$ of $M.$

There are many ways to lift a vector field from the base manifold to a
vector field in the bundle space $J^{m}(M)$ depending on the geometric
properties one wants to preserve.

One type of lift, the complete lift or prolongation, is defined by the
requirement that the tangent to the vector field $X$ in $M$ goes over to the
tangent of the vector field $W\ $in $J^{m}(M)$ at the corresponding lifted
point. Equivalently one may define the prolongation by the requirement that
under the action of the point transformation the variation of the variables
equals the difference of the derivatives before and after the action of the
one parameter transformation. For example for $\eta ^{\left[ 1\right] }~$we
have
\begin{equation}
\eta ^{\left[ 1\right] }\equiv \lim_{\varepsilon \rightarrow 0}\left[ \frac{1%
}{\varepsilon }\left( \bar{q}^{\left( 1\right) }-q^{\left( 1\right) }\right) %
\right] =\frac{d\eta }{dt}-q^{i\left( 1\right) }\frac{d\xi }{dq^{i}}.
\end{equation}

For the $k-th$ prolongation $~\eta ^{\left[ k\right] }$ follows the
recursive formula.{\large \ }%
\begin{equation}
\eta ^{\left[ k\right] }(t,q^{i},q^{i\left( 1\right) },...,q^{i\left(
m\right) })=\frac{d\eta ^{k-1}}{dt}-q^{i\left( k\right) }\frac{d\xi }{dq^{i}}%
=\frac{d^{k}}{dq^{ik}}\left( \eta -q^{i\left( 1\right) }\xi \right)
+q^{i\left( k+1\right) }\xi .
\end{equation}

Two important observations for the prolongation $\eta ^{\left[ n\right] }$
are, (a) $\eta ^{\left[ n\right] }$ is linear in $q^{i\left( n\right) }$,
and (b) $\eta ^{\left[ n\right] }$ is a polynomial in the derivatives $%
q^{i\left( 1\right) },...,q^{i\left( n\right) }$ whose coefficients are
linear homogeneous in the functions $\xi \left( x,y\right) ,$ $\eta \left(
x,y\right) $ up to nth order partial derivatives.

Concerning the general vector $W$ on $J^{m}(M)$ one defines its components
by the requirement
\begin{equation*}
X^{\left[ m\right] }(t,q^{i},q^{i\left( 1\right) },...,q^{i\left( m\right)
})=\eta ^{\left[ k\right] }(t,q^{i},q^{i\left( 1\right) },...,q^{i\left(
m\right) })+\phi ^{m}
\end{equation*}%
where $\phi ^{m}(t,q^{i},q^{i\left( 1\right) },...,q^{i\left( m\right) })$
are some functions on $J^{m}(M)$ which will be defined by additional
requirements. For example the complete lift is defined by the requirement $%
\phi ^{i}=0.$

In the case we have a manifold with $n$ independent$~$variables~$\left\{
x^{i}:i=1..n\right\} $ and $m$ dependent variables~$\left\{
u^{A}:A=1...m\right\} ,$ we consider the one parameter point transformation
in the jet space $\{x^{i},u^{A}\}$ of the form%
\begin{equation*}
\bar{x}^{i}=\Xi ^{i}\left( x^{i},u^{A},\varepsilon \right) ~~,~~\bar{u}%
^{A}=\Phi ^{A}\left( x^{i},u^{A},\varepsilon \right) .
\end{equation*}%
The infinitesimal generator is
\begin{equation}
X=\xi ^{i}\left( x^{k},u^{A}\right) \partial _{i}+\eta ^{A}\left(
x^{k},u^{A}\right) \partial _{A}  \label{MP.01}
\end{equation}%
where%
\begin{equation*}
\xi ^{i}\left( x^{k},u^{A}\right) =\frac{\partial \Xi ^{i}\left(
x^{i},u^{A},\varepsilon \right) }{\partial \varepsilon }|_{\varepsilon
\rightarrow 0}~~,~~\eta ^{A}\left( x^{k},u^{A}\right) =\frac{\partial \Phi
\left( x^{i},u^{A},\varepsilon \right) }{\partial \varepsilon }%
|_{\varepsilon \rightarrow 0}.
\end{equation*}

In a similar way, the prolongation vector is calculated to be
\begin{equation*}
X^{\left[ n\right] }=X+\eta_{i}^{A}\partial_{u_{i}}+...+\eta_{ij..i_{n}}^{A}%
\partial_{u_{ij..i_{n}}}
\end{equation*}
where now%
\begin{equation}
\eta_{i}^{A}=D_{i}\eta^{A}-u_{,j}^{A}D_{i}\xi^{j},  \label{MP.02A}
\end{equation}%
\begin{equation}
\eta_{ij..i_{n}}^{A}=D_{i_{n}}\eta_{ij..i_{n-1}}^{A}-u_{ij..k}D_{i_{n}}\xi
^{k},  \label{MP.02}
\end{equation}
and the operator $D_{i}$ is defined as%
\begin{equation}
D_{i}=\frac{\partial}{\partial x^{i}}+u_{i}^{A}\frac{\partial}{\partial u^{A}%
}+u_{ij}^{A}\frac{\partial}{\partial u_{j}^{A}}+...+u_{ij..i_{n}}^{A}\frac{%
\partial}{\partial u_{jk..i_{n}}^{A}}.
\end{equation}

In terms of the partial derivatives of the components $\xi ^{i}\left(
x^{k},u^{A}\right) ,~\eta ^{A}\left( x^{k},u^{A}\right) $ of the generator (%
\ref{MP.01}), the first and the second extension of the symmetry vector are
given as follows%
\begin{equation}
X^{\left[ 1\right] }=X+\left( \eta _{,i}^{A}+u_{,i}^{B}\eta _{,B}^{A}-\xi
_{,i}^{j}u_{,j}^{A}-u_{,i}^{A}u_{,j}^{B}\xi _{,B}^{j}\right) \partial
_{u_{i}^{A}}
\end{equation}%
\begin{equation}
X^{\left[ 2\right] }=X^{\left[ 1\right] }+\left[
\begin{array}{c}
\eta _{,ij}^{A}+2\eta _{,B(i}^{A}u_{,j)}^{B}-\xi _{,ij}^{k}u_{,k}^{A}+\eta
_{,BC}^{A}u_{,i}^{B}u_{,j}^{C}-2\xi _{,(i\left\vert B\right\vert
}^{k}u_{j)}^{B}u_{,k}^{A}+ \\
-\xi _{,BC}^{k}u_{,i}^{B}u_{,j}^{A}u_{,k}^{A}+\eta _{,B}^{A}u_{,ij}^{B}-2\xi
_{,(j}^{k}u_{,i)k}^{A}+-\xi _{,B}^{k}\left(
u_{,k}^{A}u_{,ij}^{B}+2u_{(,j}^{B}u_{,i)k}^{A}\right)%
\end{array}%
\right] \partial _{u_{ij}}.
\end{equation}%
Similar expressions can be considered for the general vector field in the
jet space $\{x^{i},u^{A}\}.$


\subsection{Invariance of functions}

A differentiable function $F\left( q^{i},q^{i(m)}\right) $ on $J^{m}(M)$ \
is said to be invariant under the action of $X$ iff

\begin{equation}
X\left( F\right) =0,  \label{InF.01}
\end{equation}%
or, equivalently, iff there exists a function $\lambda $ such that%
\begin{equation}
X\left( F\right) =\lambda F~, mod F=0.
\end{equation}

In order to determine the invariant functions of a given infinitesimal point
generator $X$ one has to solve the associated Lagrange system%
\begin{equation}
\frac{dt}{\xi }=\frac{dq^{i}}{\eta ^{i}}=...=\frac{dq^{i(m)}}{\eta ^{i(m)}}.
\end{equation}

The characteristic function or zero order invariant $w$ of the vector $X$ is
defined as follows
\begin{equation}
dw=\frac{dt}{\xi }-\frac{dq^{i}}{\eta ^{i}}.
\end{equation}%
The zero order invariant is indeed invariant under the action of $X,$ that
is $X\left( w\right) =0.$ Therefore, any function of the form $F=F\left(
w\right) $ satisfies the symmetry condition (\ref{InF.01}) and it is
invariant under the action of $X.$ The higher order invariants $v_{(b)}$ are
given by the expression
\begin{equation}
v_{(m)}=\frac{dt}{\xi }-\frac{dq^{i(m)}}{\eta ^{i(m)}}.
\end{equation}

\section{Symmetries of differential equations}

Consider an $mth$ order system of differential equations defined on $%
J^{m}(M) $ of the form $q^{a[m]}=F(t,q,...,q^{[m]}).$ We say that the point
transformation (\ref{PointTrans.1}) in $J^{m}(M)$ generated by the vector
field $X=\xi \frac{\partial }{\partial t}+\eta ^{a[A]}\frac{\partial
^{\lbrack A]}}{\partial q^{a[A]}}$ is a symmetry of the system of equations
if it leaves the \emph{set} of solutions of the system the same.
Equivalently if we consider the function $G=q^{a[m]}-F(t,q,...,q^{[m]})=0$
on $J^{m}(M)$ then a symmetry of the differential equation is a vector field
leaving $G$ invariant.

The main reason for studying the symmetries of a system of differential
equations is to find first integrals and/or invariant solutions. Both these
items facilitate the solution and the geometric / physical interpretation of
the system of equations.

In the following we shall be interested in systems of second order
differential equations (SODE) of the form $\ddot{q}^{a}-K^{a}(t,q,\dot{q}%
)=0, $ therefore we shall work on the jet space $J^{1}(M)$ which is
essentially the space $R\times TM$. In this case the infinitesimal point
transformation (\ref{PointTrans.1}) is written
\begin{eqnarray}
t^{\prime } &=&t+\varepsilon \xi +O^{2}\left( \varepsilon ^{2}\right) +\cdots
\notag \\
q^{i^{\prime }} &=&q^{i}+\varepsilon \eta ^{i}+O^{2}\left( \varepsilon
^{2}\right) +\cdots  \notag \\
\dot{q}^{i^{\prime }} &=&\dot{q}^{i}+\varepsilon \left( \dot{\eta}^{i}-\dot{%
\xi}\dot{q}^{i}+\phi ^{i}\right) +O\left( \varepsilon ^{2}\right) +\cdots
\label{TransfHol}
\end{eqnarray}%
and it is generated by the vector field
\begin{equation}
X^{W}=X^{\left[ 1\right] }+\phi ^{i}\frac{\partial }{\partial \dot{q}^{i}}
\label{SymHol}
\end{equation}%
where $\phi ^{i}(t,q,\dot{q})$ are some general (smooth) functions and $X^{%
\left[ 1\right] }$ is the prolonged vector field:%
\begin{equation}
X^{\left[ 1\right] }=\xi (t,q,\dot{q})\frac{\partial }{\partial t}+\eta
^{a}(t,q,\dot{q})\frac{\partial }{\partial q^{a}}+X^{[1]a}\frac{\partial }{%
\partial \dot{q}^{i}}  \label{SymHol1}
\end{equation}%
where
\begin{equation}
X^{[1]a}=\frac{d\eta ^{a}}{dt}-\dot{q}^{a}\frac{d\xi }{dt}.
\label{SymHol1.1}
\end{equation}%
If $\xi (t,q),\eta ^{i}(t,q)$ then $X=\xi \frac{\partial }{\partial t}+\eta
^{a}\frac{\partial }{\partial q^{a}}$ is defined on the base manifold $M$
and $X^{\left[ 1\right] }$ is called the first prolongation of $X$ in $TM.$

\section{The conservative holonomic dynamical system}

Consider the conservative holonomic system (CHS) with Lagrangian $L=\frac{1}{%
2}\gamma _{ij}\dot{q}^{i}\dot{q}^{j}-V(t,q),$ where $V(t,q)$ is the
potential of all conservative forces, whose equations of motion are%
\begin{equation}
\frac{d}{dt}\left( \frac{\partial L}{\partial \dot{q}^{i}}\right) -\frac{%
\partial L}{\partial q^{i}}=0.  \label{GenHolonEuler-Lagrange}
\end{equation}%
These are written
\begin{equation}
E^{i}(L)=0  \label{Eqnmot.1}
\end{equation}%
where $E^{i}$ is the Euler vector field in the jet space $J^{1}(t,q,\dot{q})$
\begin{equation*}
E_{i}=\frac{d}{dt}\frac{\partial }{\partial \dot{q}^{i}}-\frac{\partial }{%
\partial q^{i}}.
\end{equation*}%
\ Replacing the Lagrangian in (\ref{GenHolonEuler-Lagrange}) we find
\begin{equation}
\ddot{q}^{i}=\omega ^{i}  \label{Eqnmotion.2}
\end{equation}%
where
\begin{equation}
\omega ^{i}(t,q,\dot{q})=-V^{,i}-\Gamma _{jk}^{i}\dot{q}^{j}\dot{q}^{k}
\label{Se.1.1}
\end{equation}%
The CHS defines two important geometric quantities in the jet space $%
J^{1}(t,q,\dot{q})$

a. The kinetic metric $\gamma _{ij}=\frac{\partial ^{2}L}{\partial
q^{i}\partial q^{j}}$ (we assume the Lagrangian to be regular, that is $\det
\frac{\partial ^{2}L}{\partial q^{i}\partial q^{j}}\neq 0$ so that the
kinetic metric is non-degenerate) which is essentially the kinetic energy of
the dynamical system. This metric is different from the metric of the space
where motion occurs. It is a positive definite metric of dimension depending
on the degrees of freedom of the dynamical system.

b. The Hamiltonian vector field $\Gamma $
\begin{equation}
\Gamma =\frac{d}{dt}=\frac{\partial }{\partial t}+\frac{\partial }{\partial
q^{i}}\dot{q}^{i}+\omega ^{i}\frac{\partial }{\partial \dot{q}^{i}}.
\label{Se.1}
\end{equation}

Obviously the Hamiltonian vector field is characteristic of the dynamical
equations (\ref{Eqnmotion.2}) and can be defined in all cases irrespective
of the Lagrangian function.

In the following we shall restrict our considerations to the symmetries of
second order differential equations of the form (\ref{Eqnmotion.2}).

\section{Types of symmetries}

There are various types of multi parameter point transformations in $%
J^{1}(M) $ which generate a symmetry of a system of second order
differential equations (SODE). In cosmology - at least at the current status
- it appears that two of them are of importance, that is, the Lie symmetries
and the Noether symmetries. Both type of symmetries lead to first integrals
hence providing ways to solve the considered cosmological equations and
shall be discussed in the following. In case the components $\xi ,\eta ^{a}$
of the generators are functions of $t,q^{a}$ only the symmetries are called
point symmetries.

The requirements for each type of symmetry lead to a set of conditions which
when solved give the generators of the corresponding point transformation
and consequently the way to determine first integrals. As it has been shown
the generators of these symmetries for autonomous conservative holonomic
systems are related to the collineations of the kinetic metric. Specifically
it has been shown{ \cite{p1}} that the Lie point symmetries are
elements of the special projective algebra and the Noether point symmetries
elements of the homothetic algebra \cite{p2}. Concerning the partial
differential equations the generators are related with the conformal group
of the kinetic metric{\ \cite{p3,p4}} .

\subsection{Lie symmetries}

\textbf{Definition:} The vector field $X^{w}$ on the jet space $J^{1}(M)$ is
a (dynamical) Lie symmetry of the equations (\ref{Eqnmotion.2}) if it is a
symmetry of the Hamiltonian vector field; that is if the following condition
is satisfied%
\begin{equation}
L_{X^{W}}\Gamma =\lambda \Gamma  \label{SymHolCond}
\end{equation}%
where $\lambda (t,q,\dot{q})$ is a function to be defined.

Geometrically a Lie symmetry preserves the set of solutions of the equations
(\ref{Eqnmotion.2}), in the sense, that under the action of the point
transformation generated by $X^{W}$ a point of a solution curve is
transformed to a point of another solution curve of (\ref{Eqnmotion.2}).

The symmetry condition (\ref{SymHolCond}) gives:%
\begin{equation}
\left[ X^{W},\Gamma \right] =-\Gamma \left( \xi \right) \frac{\partial }{%
\partial t}+\left( X^{W}\left( \dot{q}^{a}\right) -\Gamma \left( \eta
^{a}\right) \right) \frac{\partial }{\partial q^{a}}+\left( X^{W}\left(
F^{a}\right) -\Gamma \left( X^{a}+\phi ^{a}\right) \right) \frac{\partial }{%
\partial \dot{q}^{a}}=\lambda \Gamma .  \label{SymHolCond.1}
\end{equation}%
Condition (\ref{SymHolCond.1}) is equivalent to the following system of
equations:%
\begin{eqnarray}
-\Gamma \left( \xi \right) &=&\lambda  \notag \\
\Gamma \left( \eta ^{a}\right) -\Gamma \left( \xi \right) \dot{q}^{a}-\phi
^{a} &=&X^{a}  \label{LieCon2} \\
\xi \frac{\partial \omega ^{a}}{\partial t}+\eta ^{b}\frac{\partial \omega
^{a}}{\partial q^{b}}+\left( X^{b}+\phi ^{b}\right) \frac{\partial \omega
^{a}}{\partial \dot{q}^{b}}+\Gamma \left( \xi \right) F^{a} &=&\Gamma \left(
X^{a}+\phi ^{a}\right) .  \label{LieCon3}
\end{eqnarray}

The second condition gives $\phi ^{a}=0$ so that $X^{W}=X^{[1]}$ where
\begin{equation*}
X^{[1]}=\xi \left( t,q^{k},\dot{q}^{k}\right) \partial _{t}+\eta ^{a}\left(
t,q^{k},\dot{q}^{k}\right) \partial _{a}+\left[ \Gamma \left( \eta
^{a}\right) -\Gamma \left( \xi \right) \dot{q}^{a}\right] \frac{\partial }{%
\partial \dot{q}^{a}}
\end{equation*}%
is given by (\ref{SymHol1}). Then the third condition (\ref{LieCon3}) becomes%
\begin{equation*}
\xi \frac{\partial \omega ^{a}}{\partial t}+\eta ^{b}\frac{\partial \omega
^{a}}{\partial q^{b}}+\left( \Gamma \left( \eta ^{a}\right) -\Gamma \left(
\xi \right) \dot{q}^{a}\right) \frac{\partial \omega ^{a}}{\partial \dot{q}%
^{b}}+\Gamma \left( \xi \right) \omega ^{a}=\Gamma \left( \Gamma \left( \eta
^{a}\right) -\Gamma \left( \xi \right) \dot{q}^{a}\right)
\end{equation*}%
and can be written more compactly as
\begin{equation}
X^{[1]}(\omega ^{a})+\Gamma \left( \xi \right) \omega ^{a}=\Gamma \left(
\Gamma \left( \eta ^{a}\right) -\Gamma \left( \xi \right) \dot{q}^{a}\right)
.  \label{LieCon4}
\end{equation}

We note that the functions $\phi ^{a}$ do not take part into the conditions
of Lie symmetries and can be omitted. The exact form of the Lie symmetry
conditions depends on the functional dependence of the functions $\xi (t,q,%
\dot{q}),\eta ^{a}(t,q,\dot{q})$ and of course on the form of the
Hamiltonian field. In the following we consider two important cases of Lie
symmetries.

\subsection{Lie point symmetries}

In case $\xi (t,q),\eta ^{a}(t,q)$ the Lie symmetry is called a Lie point
symmetry. For the special class of differential equations of the form%
\begin{equation}
\ddot{q}+\Gamma _{jk}^{i}\dot{q}^{j}\dot{q}^{k}+V^{,i}\left( t,q^{k}\right)
=0
\end{equation}%
the Lie symmetry condition (\ref{LieCon3}) leads to the following system of
covariant conditions {\ \cite{Tsam10}}

\begin{equation}
L_{\eta }V^{;a}+\eta _{,tt}^{a}+2V^{;a}\xi _{,t}+\xi V_{,t}^{;a}=0
\label{NBH1}
\end{equation}%
\begin{equation}
2\eta _{;b|t}^{a}-\delta _{b}^{a}\xi _{,tt}+\left( 2\delta _{c}^{a}\xi
_{;b}+\delta _{b}^{a}\xi _{;c}\right) V^{;c}=0  \label{NBH2}
\end{equation}%
\begin{equation}
\mathcal{L}_{\eta }\Gamma _{bc}^{a}=2\delta _{(b}^{a}\xi _{,c)t}
\label{NBH3}
\end{equation}%
\begin{equation}
\delta _{(d}^{a}\xi _{;bc)}=0.  \label{NBH4}
\end{equation}

The use of an algebraic computing program does not reveal directly the Lie
symmetry conditions in this geometric form. Equation (\ref{NBH4}) implies
that $\xi _{,a}$ is a gradient KV of the kinetic metric. Equation (\ref{NBH3}%
) means that $\eta ^{i}$ is a special projective collineation of the metric
with projective function $\xi _{,t}.$ The remaining two equations (\ref{NBH1}%
)and (\ref{NBH2}) are constraint conditions, which relate the components $%
\xi ,n^{i}$ of the Lie point symmetry vector with the potential function $%
V(t,q).$

Conditions (\ref{NBH1}) - (\ref{NBH4}) can be obtained as special cases from
known results. Indeed in \cite{Tsam10} it has been shown that the conditions
for the point Lie symmetry of the dynamical equations
\begin{equation}
\ddot{q}^{a}+\Gamma _{jk}^{i}\dot{q}^{j}\dot{q}^{k}+P^{i}\left(
t,q^{k}\right) =0  \label{Duk.1}
\end{equation}%
are the following:%
\begin{align}
L_{\eta }P^{i}+2\xi ,_{t}P^{i}+\xi P^{i},_{t}+\eta ^{i},_{tt}& =0
\label{de.13} \\
\left( \xi ,_{k}\delta _{j}^{i}+2\xi ,_{j}\delta _{k}^{i}\right) P^{k}+2\eta
^{i},_{t|j}-\xi ,_{tt}\delta _{j}^{i}& =0  \label{de.14} \\
L_{\eta }\Gamma _{jk}^{i}-2\xi ,_{t(j}\delta _{k)}^{i}& =0  \label{de.15} \\
\xi _{(,j|k}\delta _{d)}^{i}& =0.  \label{de.16}
\end{align}%
In order to obtain (\ref{NBH1}) - (\ref{NBH4}) one simply replaces in (\ref%
{de.13}) - (\ref{de.16}) $P^{i}=V^{,i}\left( t,q^{k}\right) $.

\subsection{Lie point symmetries and first integrals}

It is possible that a Lie point symmetry leads to a first integral. These
integrals have been called Hojman integrals and belong to the class of
non-Noetherian first integrals \cite{hoj1}. As shall be discussed below by
means of the Inverse Noether Theorem one is able to associate a Noether
symmetry to a given quadratic first integral. In this sense as far as the
first integrals are concerned, Noether symmetries are the prevailing ones. \
Concerning the Hojman symmetries we have the following \cite{hoj1}

\textbf{Proposition:} i. Necessary and sufficient condition that the point
transformation $q^{i^{\prime }}=q^{i}+\varepsilon \eta ^{i}\left( t,q^{k},%
\dot{q}^{k}\right) $ is a Lie point symmetry of the (SODE) $\ddot{q}%
^{j}-\omega ^{j}\left( t,q^{i},\dot{q}^{i}\right) =0$ is that the generator $%
\eta ^{i}$ satisfies the conditions%
\begin{eqnarray}
\Gamma (\Gamma \eta ^{i})-\eta ^{\left[ 1\right] }\left( \omega ^{j}\right)
&=&0\text{ or}  \notag \\
\ddot{\eta}^{j}-\eta ^{\lbrack 1]}(\omega ^{j}) &=&0  \label{HojLie.0.1}
\end{eqnarray}%
where $\ddot{\eta}^{j}=\Gamma (\Gamma (\eta ^{j})$ and $\eta ^{\lbrack
1]}=\eta ^{i}\frac{\partial }{\partial q^{i}}+\dot{\eta}^{i}\frac{\partial }{%
\partial \dot{q}^{i}}.$

ii. The scalar
\begin{equation*}
I_{2}=\frac{1}{\gamma }\frac{\partial }{\partial q^{j}}\left( \gamma \eta
^{j}\right) +\frac{\partial \dot{\eta}^{j}}{\partial \dot{q}^{j}}
\end{equation*}%
is a first integral of the ODE $\ddot{q}^{i}=\omega ^{i}$ iff

a. The vector $\eta ^{i}$ is a Lie symmetry of the SODE $\ddot{q}^{i}=\omega
^{i}$

b. The function $\gamma (t,q,\dot{q})$ is defined by the condition
\begin{equation*}
Trace\left( \frac{\partial \omega ^{i}}{\partial \dot{q}^{j}}\right) =\frac{%
\partial \omega ^{j}}{\partial \dot{q}^{j}}=-\frac{d}{dt}\ln \gamma \left(
q^{k}\right) .
\end{equation*}

One solution for all $\eta ^{i}$ is $\omega ^{j}=a$ $\dot{q}^{j}+\omega
^{j}(t,q^{i})$ where $a=$const$.$ For this solution we have that the SODE\
has the generic form:%
\begin{equation*}
\ddot{q}^{j}-a\dot{q}^{j}=\omega ^{j}(t,q^{i}).
\end{equation*}%
which is the equation for forced motion with with linear dumping $a.$

\subsection{Noether point symmetries}

Noether point symmetries concern Lagrangian dynamical systems and are
defined as follows.

\textbf{Definition:} Suppose that $A\left( q^{i}\dot{q}^{i}\right) $ is the
functional (the action integral)%
\begin{equation}
A\left( q^{i},\dot{q}^{i}\right) =\int_{t_{1}}^{t_{2}}L\left( t,q^{i},\dot{q}%
^{i}\right) dt.  \label{FuncAction}
\end{equation}%
The vector field $X^{W}=X^{\left[ 1\right] }+\phi ^{i}\frac{\partial }{%
\partial \dot{q}^{i}}$ in the jet space $J^{1}(M)$ generating the point
transformation (\ref{TransfHol}) is said to be a Noether symmetry of the
dynamical system with Lagrangian $L$ if

a. The action integral under the action of the point transformation
transforms as follows%
\begin{equation}
A^{\prime }\left( q^{i^{\prime }},\dot{q}^{i^{\prime }}\right) =A\left(
q^{i},\dot{q}^{i}\right) +\epsilon \int_{t_{1}}^{t_{2}}\frac{df(t,q^{i},\dot{%
q}^{i})}{dt}dt  \label{FuncActionTransf}
\end{equation}%
where $f(t,\epsilon )$ is a smooth function.

b. The infinitesimal transformation causes a zero end point variation (i.e.
the end points of the integral remain fixed)$.$

Noether symmetries use the fact that when we add a perfect differential to a
Lagrangian the equations of motion do not change. Hence the set of solutions
remains the same. This is another view of the "invariance" of the dynamical
equations (\ref{Eqnmotion.2}).

The condition for a Noether symmetry under the action of the point
transformation (\ref{TransfHol}) is

\begin{equation}
X^{\left[ 1\right] }\left( L\right) +\phi ^{i}\frac{\partial L}{\partial
\dot{q}^{i}}+L\left( t,q^{i},\dot{q}^{i}\right) \dot{\xi}=\dot{f}
\label{GenHolonNoether's1Cond}
\end{equation}%
Eqn. (\ref{GenHolonNoether's1Cond}) is called the weak Noether condition and
it is also known as the First Noether theorem.

\subsection{First integral defined by a Noether symmetry}

Next we determine the conditions which a Noether symmetry must satisfy in
order to lead to a first integral of Lagrange equations. Expanding the
Noether condition (\ref{GenHolonNoether's1Cond}) and making use of Lagrange
equations (\ref{Eqnmotion.2}) we find
\begin{equation}
\phi ^{i}\frac{\partial L}{\partial \dot{q}^{i}}=\frac{d}{dt}\left( f-L\xi -%
\frac{\partial L}{\partial \dot{q}^{i}}\left( \eta ^{i}-\xi \dot{q}%
^{i}\right) \right) .  \label{GenHolonNoether'2 Cond}
\end{equation}%
This leads to what is known as the Second Noether Theorem.

\textbf{Proposition:} The quantity%
\begin{equation}
I=f-L\xi -\frac{\partial L}{\partial \dot{q}^{i}}\left( \eta ^{i}-\xi \dot{q}%
^{i}\right)  \label{GenHolonNoether's1Integr}
\end{equation}%
is a first integral for the holonomic system defined by (\ref%
{GenHolonEuler-Lagrange}) provided the functions $\phi ^{i}$ (i.e. the
variations along the fibers) vanish. In this case the weak Noether condition
(\ref{GenHolonNoether's1Cond}) becomes%
\begin{equation}
X^{\left[ 1\right] }\left( L\right) +L\left( t,q^{i},\dot{q}^{i}\right) \dot{%
\xi}=\dot{f}  \label{GenHolonNoether'sFinCond}
\end{equation}%
and the Noether symmetry reduces to a special Lie symmetry (that is a Lie
symmetry which in addition satisfies the Noether condition). The function $f$
is called the Noether or the gauge function.

We remark that contrary to what is generally believed a Noether point
symmetry for an autonomous conservative holonomic system does not lead
necessarily to a first integral of the equations of motion. Indeed the weak
Noether symmetry condition (\ref{GenHolonNoether's1Cond}) is more general
than the standard Noether condition (\ref{GenHolonNoether'sFinCond}) because
it holds for general $\phi ^{a}$ whereas the latter holds only for $\phi
^{a}=0.$

In the following by a Noether symmetry we shall mean a point Noether
symmetry which leads to a first integral, that is the quantities $\phi
^{a}=0.$ These Noether symmetries are special Lie symmetries which satisfy
condition (\ref{GenHolonNoether'sFinCond}). It has been remarked already
above \cite{p2} the Lie point symmetries are the elements of the special
projective collineations of the kinetic metric and the Lie point symmetries
which are Noether point symmetries are elements of the homothetic subalgebra.

\section{Generalized Killing equations}

We decompose the Noether condition along the vector $\frac{d}{dt}$ and
normal to It. In order to do that we expand the overdot terms and assume
that $\ddot{q}^{i}$ are independent variables. The lhs is:%
\begin{eqnarray*}
&&L\left( \frac{\partial \xi }{\partial t}+\dot{q}^{i}\frac{\partial \xi }{%
\partial q^{i}}+\ddot{q}^{i}\frac{\partial \xi }{\partial \dot{q}^{i}}%
\right) +\xi \frac{\partial L}{\partial t}+\eta ^{i}\frac{\partial L}{%
\partial q^{i}} \\
&&+\left( \frac{\partial \eta ^{i}}{\partial t}+\dot{q}^{j}\frac{\partial
\eta ^{i}}{\partial q^{j}}+\ddot{q}^{j}\frac{\partial \eta ^{i}}{\partial
\dot{q}^{j}}-\dot{q}^{i}\frac{\partial \xi }{\partial t}-\dot{q}^{i}\dot{q}%
^{j}\frac{\partial \xi }{\partial q^{j}}-\dot{q}^{i}\ddot{q}^{j}\frac{%
\partial \xi }{\partial \dot{q}^{j}}\right) \frac{\partial L}{\partial \dot{q%
}^{i}} \\
&=&L\left( \frac{\partial \xi }{\partial t}+\dot{q}^{i}\frac{\partial \xi }{%
\partial q^{i}}\right) +\xi \frac{\partial L}{\partial t}+\eta ^{i}\frac{%
\partial L}{\partial q^{i}}+\left( \frac{\partial \eta ^{i}}{\partial t}+%
\dot{q}^{j}\frac{\partial \eta ^{i}}{\partial q^{j}}-\dot{q}^{i}\frac{%
\partial \xi }{\partial t}-\dot{q}^{i}\dot{q}^{j}\frac{\partial \xi }{%
\partial q^{j}}\right) \frac{\partial L}{\partial \dot{q}^{i}} \\
&&+\ddot{q}^{j}\left( \frac{\partial \xi }{\partial \dot{q}^{j}}L+\left(
\frac{\partial \eta ^{i}}{\partial \dot{q}^{j}}-\dot{q}^{i}\frac{\partial
\xi }{\partial \dot{q}^{j}}\right) \frac{\partial L}{\partial \dot{q}^{i}}%
\right)
\end{eqnarray*}%
the rhs gives:%
\begin{equation*}
\frac{df}{dt}=\frac{\partial f}{\partial t}+\dot{q}^{i}\frac{\partial f}{%
\partial q^{i}}+\ddot{q}^{i}\frac{\partial f}{\partial \dot{q}^{i}}.
\end{equation*}%
Therefore we obtain the following equivalent system of the equations:%
\begin{eqnarray}
L\left( \frac{\partial \xi }{\partial t}+\dot{q}^{i}\frac{\partial \xi }{%
\partial q^{i}}\right) +\xi \frac{\partial L}{\partial t}+\eta ^{i}\frac{%
\partial L}{\partial q^{i}}+\left( \frac{\partial \eta ^{i}}{\partial t}+%
\dot{q}^{j}\frac{\partial \eta ^{i}}{\partial q^{j}}-\dot{q}^{i}\frac{%
\partial \xi }{\partial t}-\dot{q}^{i}\dot{q}^{j}\frac{\partial \xi }{%
\partial q^{j}}\right) \frac{\partial L}{\partial \dot{q}^{i}} &=&\frac{%
\partial f}{\partial t}+\dot{q}^{i}\frac{\partial f}{\partial q^{i}}
\label{7} \\
\frac{\partial \xi }{\partial \dot{q}^{j}}L+\left( \frac{\partial \eta ^{i}}{%
\partial \dot{q}^{j}}-\dot{q}^{i}\frac{\partial \xi }{\partial \dot{q}^{j}}%
\right) \frac{\partial L}{\partial \dot{q}^{i}} &=&\frac{\partial f}{%
\partial \dot{q}^{j}}.  \label{8}
\end{eqnarray}%
These equations have been called the generalized Killing equations (see eqns
(17),(18) of Djukic in \cite{d2}).

We note that the generalized Killing equations have $2n+2$ unknowns (the $%
\xi ,\eta ^{i},f,\phi ^{i})$ and are only $n+1$ equations. Therefore there
is not a unique solution $X^{W}$ and we are free to fix $n+1$ variables in
order to get a solution. However this is not a problem because all these
solutions admit the same first integral $I$\ of the dynamical equations
(because all satisfy the Noether condition (\ref{GenHolonNoether'sFinCond})).

\subsection{How to solve the generalized Killing equations}

Suppose that by some method we have determined a quadratic first integral $I$
of the dynamical equations. Our purpose is to determine a gauged Noether
symmetry which will admit the given quadratic first integral. Assume the $%
n+1 $ gauge conditions $\xi =0,\phi ^{i}=0.$ Suppose $L(t,q^{i},\dot{q}^{i})$
is the Lagrangian part of the dynamical equations\footnote{%
We only need the kinetic energy which will define the kinetic metric.}. Let $%
X^{W}=\eta ^{i}(t,q^{i},\dot{q}^{i})\frac{\partial }{\partial q^{i}}+\dot{%
\eta}^{i}\frac{\partial }{\partial \dot{q}^{i}}$ be the vector field
generating a Noether symmetry which admits the first integral $I=f-\left(
\eta ^{i}-\xi \dot{q}^{i}\right) \frac{\partial L}{\partial \dot{q}^{i}}%
-L\xi $ where $f(t,q^{i},\dot{q}^{i})$ is the Noether gauge function. We
compute%
\begin{equation*}
\frac{\partial I}{\partial \dot{q}^{i}}=\frac{\partial f}{\partial \dot{q}%
^{i}}-\frac{\partial \eta ^{j}}{\partial \dot{q}^{i}}\frac{\partial L}{%
\partial \dot{q}^{j}}+\gamma _{ij}\eta ^{j}
\end{equation*}%
where $\gamma _{ij}=\frac{\partial ^{2}L}{\partial \dot{q}^{i}\partial \dot{q%
}^{j}}$is the kinetic metric determined by the Lagrangian $L.$ Then the
second equation \emph{(\ref{8}) }gives that $\frac{\partial f}{\partial \dot{%
q}^{i}}-\frac{\partial \eta ^{j}}{\partial \dot{q}^{i}}\frac{\partial L}{%
\partial \dot{q}^{j}}=0\ $ $\ $and we find eventually the expression%
\begin{equation}
\eta ^{i}=\gamma ^{ij}\frac{\partial I}{\partial \dot{q}^{i}}.
\label{CartanCondition}
\end{equation}

The vector $X^{W}=\eta ^{i}\frac{\partial }{\partial q^{i}}$ $+\dot{\eta}^{i}%
\frac{\partial }{\partial \dot{q}^{i}}$we have determined is not the only
one possible. For example one may specify the gauge function $f$ and assume
a form for $\xi (t,q)$ (while maintain the gauge $\phi ^{i}=0)$ and then use
equation (\ref{8}) to determine the solution $X^{W}$ (see example below).
However in all cases the first integral $I\ $is the same.

The first integral of a Noether point symmetry $X^{W}=X^{[1]}$ is in
addition an invariant of the Noether generator, that is
\begin{equation}
\mathbf{X}^{\left[ 1\right] }(I)=0  \label{EqnNoether.3a1}
\end{equation}%
The result (\ref{EqnNoether.3a1}) means that a Noether point symmetry
results in a twofold reduction of the order of Lagrange equations resulting
from the given action integral. This is done as follows. The first integral $%
I(t,q^{i},\dot{q}^{i})$ can be used to replace one of the second order
equations by the first order ODE $I(t,q^{i},\dot{q}^{i})=I_{0},$ where $%
I_{0} $ is a constant fixed by the initial (or boundary) conditions.
Property (\ref{EqnNoether.3a1}) says that this new equation admits the Lie
symmetry $X^{W}$ (because $X^{W}(I-I_{0})=0$) therefore it can be used to
integrate the equation once more, according to well known methods.

Noether symmetries are mainly applied to construct first integrals which are
important to determine the solution of a given dynamical system. It is
possible that there exist different (i.e. not differing by a perfect
differential) Lagrangians describing the same dynamical equations. These
Lagrangians have different Noether symmetries (see \cite{jlm1,jlm2,jlm3}).
Therefore it is clear that when we refer to a Noether symmetry of a given
dynamical system we should mention always the Lagrangian function we are
assuming.

\bigskip\ Consider the Emden - Fauler equation%
\begin{equation*}
t\ddot{q}+2\dot{q}+at^{\nu }q^{2\nu +3}=0
\end{equation*}%
where $a,\nu $ are arbitrary constants. This equation defines a conservative
holonomic dynamical system with Lagrangian%
\begin{equation*}
L=\frac{1}{2}\left( t^{2}\dot{q}^{2}-\frac{a}{\nu +2}t^{\nu +1}q^{2\nu
+4}\right) .
\end{equation*}

Assume now that the function $f(t,q,\dot{q})=-AtL$ \ where $A$ is some
constant and assume further that $\xi (t,q).$Eqn. (\ref{8}) \ gives%
\begin{equation*}
\frac{\partial \eta }{\partial \dot{q}}\frac{\partial L}{\partial \dot{q}}%
=-At\frac{\partial L}{\partial \dot{q}}\Rightarrow \eta =-At\dot{q}+F(t,q)
\end{equation*}%
where $F$ is an arbitrary function of its arguments. Replacing in (\ref{7})
we find
\begin{equation*}
L\left( \frac{\partial \xi }{\partial t}+\dot{q}\frac{\partial \xi }{%
\partial q}\right) +\eta \frac{\partial L}{\partial q}+\xi \frac{\partial L}{%
\partial t}+\left( \frac{\partial \eta }{\partial t}+\dot{q}\frac{\partial
\eta }{\partial q}-\dot{q}\frac{\partial \xi }{\partial t}-\dot{q}^{2}\frac{%
\partial \xi }{\partial q}\right) \frac{\partial L}{\partial \dot{q}}=-\frac{%
\partial f}{\partial t}+\dot{q}\frac{\partial f}{\partial q}
\end{equation*}%
which provides the following system%
\begin{eqnarray*}
\xi &=&\xi \left( t\right) \\
\left( \frac{\xi }{t}-\frac{1}{2}\xi _{,t}\right) +F_{,q} &=&-\frac{1}{2}A \\
F &=&F\left( q\right) \\
\frac{1}{2\left( \nu +2\right) }\left( \xi _{,t}+\frac{\xi }{t}\left( \nu
+1\right) \right) +\frac{F}{q} &=&-\frac{1}{2}A.
\end{eqnarray*}

The solution of the latter system is
\begin{equation}
\xi =-\left( 2c_{0}+A\right) t~,~\eta =-At\dot{q}+c_{0}q.
\end{equation}

Still we do not know the parameters $A,c_{0}.$ In order to compute them we
turn to the first integral $I=f-\left( \eta ^{i}-\xi \dot{q}^{i}\right)
\frac{\partial L}{\partial \dot{q}^{i}}-L\xi .$ Replacing we have:%
\begin{eqnarray*}
I &=&-\frac{1}{2}A\left( t^{2}\dot{q}^{2}-\frac{a}{\nu +2}t^{\nu +1}q^{2\nu
+4}\right) t-\left( -At\dot{q}+c_{0}q+\left( 2c_{0}+A\right) t\dot{q}\right)
t^{2}\dot{q}+\frac{1}{2}\left( t^{2}\dot{q}^{2}-\frac{a}{\nu +2}t^{\nu
+1}q^{2\nu +4}\right) \left( 2c_{0}+A\right) t \\
&=&-c_{0}\left( t^{2}q\dot{q}+t^{3}\dot{q}^{2}+\frac{a}{\nu +2}t^{\nu
+2}q^{2\left( \nu +2\right) }\right)
\end{eqnarray*}%
Hence:\emph{\ }%
\begin{equation*}
I=t^{3}\dot{q}^{2}+t^{2}q\dot{q}+\frac{a}{2+\nu }t^{\nu +2}q^{2(\nu +2)}=%
\text{const.}
\end{equation*}%
Having computed $I$ we compute $\eta ^{i}$ from the relation $\eta
^{i}=\gamma ^{ij}\frac{\partial I}{\partial \dot{q}^{i}}.$ The $\gamma
_{ij}=t^{2}$, that is, $\gamma ^{ij}=\frac{1}{t^{2}}.$ Then $\eta =\frac{1}{%
t^{2}}(2t^{3}\dot{q}+t^{2}q)=2t\dot{q}+q.$ Comparing this with what we have
found we get $A=-2,$ $c_{0}=1.$ We note that for these values of $A,c_{0}$
the $\xi =0$ as it is correct because the relation $\eta ^{i}=\gamma ^{ij}%
\frac{\partial I}{\partial \dot{q}^{i}}$ is valid only under the assumption $%
\xi =0.$

\section{The Inverse Noether Theorem}

One question which arises concerns the extend that the first integrals
provided by different types of symmetry of a system of differential
equations are independent. In this section we show that to any quadratic
first integral one may associate a Noether symmetry which provides that
integral as a Noether integral.

Suppose we\ have a quadratic first integral $I$ of a Lagrangian system with
non-degenerate kinetic metric. We\ define a vector $\eta ^{i}(t,q,\dot{q})$
and a function $f(t,q,\dot{q})$ by the requirement%
\begin{equation*}
I=f-\eta ^{i}\dot{q}^{i}.
\end{equation*}%
Because $I$ is quadratic in the velocities $\eta ^{i}$ must be linear in the
velocities and $f$ must be at most quadratic in the velocities. We\ choose $%
\eta _{i}=a_{i}(t,q)+b_{ij}(t,q)\dot{q}^{j}$ and $f=\frac{1}{2}c_{ij}(t,q)%
\dot{q}^{i}\dot{q}^{j}+d_{i}(t,q)\dot{q}^{i}+e_{i}(t,q).$ Then we\ have%
\begin{equation*}
\eta _{i}=-\frac{\partial I}{\partial \dot{q}^{i}}+\frac{\partial f}{%
\partial \dot{q}^{i}}=-\frac{\partial I}{\partial \dot{q}^{i}}+c_{ij}(t,q)%
\dot{q}^{j}+d_{i}(t,q)
\end{equation*}%
and replacing $\eta _{i}$%
\begin{equation*}
a_{i}(t,q)+b_{ij}(t,q)\dot{q}^{i}=-\frac{\partial I}{\partial \dot{q}^{i}}%
+c_{ij}(t,q)\dot{q}^{j}+d_{i}(t,q).
\end{equation*}%
Let us assume that $I$ has the general form
\begin{equation*}
I=\frac{1}{2}A_{ij}(t,q)\dot{q}^{i}\dot{q}^{j}+B_{i}(t,q)\dot{q}^{i}+C(t,q).
\end{equation*}%
Then we have%
\begin{equation*}
a_{i}(t,q)+b_{ij}(t,q)\dot{q}^{j}=A_{ij}(t,q)\dot{q}%
^{j}+B_{i}(t,q)+c_{ij}(t,q)\dot{q}^{j}+d_{i}(t,q)
\end{equation*}%
from which follows%
\begin{eqnarray*}
b_{ij} &=&-A_{ij}+c_{ij} \\
a_{i} &=&-B_{i}+d_{i}
\end{eqnarray*}%
This system has an infinite number of solutions. To pick up one solution we
have to define $c_{ij},d_{i}.$ One choice is $c_{ij}=-A_{ij}$ which gives $%
b_{ij}=-2A_{ij}$ and $d_{i}=0$ which implies $a_{i}=-B_{i.}$ Therefore one
answer is%
\begin{eqnarray*}
\eta _{i} &=&2A_{ij}\dot{q}^{j}+B_{i} \\
f &=&\frac{1}{2}A_{ij}(t,q)\dot{q}^{i}\dot{q}^{j}+e_{i}(t,q).
\end{eqnarray*}%
Then the vector $X^{W}=\eta ^{i}\frac{\partial }{\partial q^{i}}+\eta
^{\lbrack 1]i}\frac{\partial }{\partial \dot{q}^{i}}$%
\begin{equation*}
X^{W}=\eta ^{i}\frac{\partial }{\partial q^{i}}+\eta ^{\lbrack 1]i}\frac{%
\partial }{\partial \dot{q}^{i}}
\end{equation*}%
generates a point transformation which is a gauged Noether symmetry (in the
gauge $\xi =0,\phi ^{i}=0)$ of the conservative holonomic dynamical system
admitting $I\ $as a Noether integral.

Obviously all quadratic first integrals of SODEs correspond to a Noether
symmetry which is computed as indicated above.

\section{Symmetries of SODEs in flat space}

Obviously an area where symmetries of ODEs play an important role is the
cases in which the kinetic metric (not necessarily the spacetime metric)\ is
flat. These cases cover significant part of Newtonian Physics where the
kinetic energy is a positive definite metric with constant coefficients
therefore there is always a coordinate transformation in the configuration
space which brings the metric to the Euclidian metric. Similar remarks apply
to Special Relativity and -as we shall see - to Cosmology.

The basic result in this cases is that the maximum number of Lie point
symmetries a SODE\ can have is $n(n+2)$ and the maximum number of Noether
point symmetries{\LARGE \ }$\frac{n(n+1)}{2}+1.$ Moreover, the number of
point symmetries which a SODE can posses is exactly one of 0,~1,~2,~3, or 8~%
\cite{fm1}.~ Similar results exist for higher-order differential equations
\cite{fm2}.

Lie has shown \cite{lie1} the important result that \textquotedblleft for
all the second order ordinary differential equations$\ $which are invariant
under the elements of the $sl\left( 3,R\right) $, there exists a
transformation of variables which brings the equation to the form $x^{\ast
\prime \prime }=0$ and vice versa\textquotedblright\ .

In current cosmological models of importance of interest is the case $n=2,$
therefore we shall restrict our attention to two cases

i. The case the of systems which admit the maximum number of Lie point
symmetries which for $n=2$ is eight and span the algebra $sl(3,R)$.

ii. The case that the Lie point symmetries span the algebra $sl(2,R).$

\subsection{The case of $sl(3,R)$ algebra}

The prototype dynamical system which admits the $sl(3,R)$ algebra of eight
Lie point symmetries is the Newtonian free particle moving in one dimension
whose dynamical equation is
\begin{equation}
\ddot{x}=0  \label{NS.07}
\end{equation}%
where $x=x\left( t\right) $ and a dot means differentiation with respect to
the time parameter $t$.

Let $X=\xi \left( t,x\right) \partial _{t}+\eta \left( t,x\right) \partial
_{i}$ be the generator of a Lie point symmetry of (\ref{NS.07}). The Lie
point symmetries of (\ref{NS.07}) are given by the special projective
vectors of $E^{2}$ (see condition (\ref{NBH3}) )

\begin{equation*}
X_{1}=\partial _{t}~,~X_{2}=\partial _{x}~,~X_{3}=t\partial
_{x}~,~X_{4}=t^{2}\partial _{t}+tx\partial _{x}~,~~X_{5}=t\partial _{t}~~,~
\end{equation*}%
\begin{equation*}
X_{6}=x\partial _{x}~~,~X_{7}=tx\partial _{t}+x^{2}\partial
_{x}~~,~~X_{8}=x\partial _{t}.
\end{equation*}

To show the validity of the aforementioned Lie's result we consider the
harmonic oscillator%
\begin{equation}
\ddot{x}+x=0,  \label{Ns.09}
\end{equation}%
which also admits the eight Lie point symmetries \cite{wfa1}%
\begin{equation*}
\bar{X}_{1}=\partial _{t}~,~\bar{X}_{2}=\cos t\partial _{x}~~,~\bar{X}%
_{3}=\sin t\partial _{x}~~,~\bar{X}_{4}=x\partial _{x}~,
\end{equation*}%
\begin{equation*}
\bar{X}_{5}=\sin 2t~\partial _{t}+x\cos 2t~\partial _{x}~,~\bar{X}_{6}=\cos
2t~\partial _{t}-x\sin 2t~\partial _{x}~,
\end{equation*}%
~%
\begin{equation*}
\bar{X}_{7}=x\sin t~\partial _{t}+x^{2}\cos t~\partial _{x}~,~\bar{X}%
_{8}=x\cos t\partial _{t}-x^{2}\sin t~\partial _{x},
\end{equation*}%
which form another basis of\ the $Sl\left( 3,R\right) $ Lie algebra. The
transformation which relates the two different representations of the $%
Sl\left( 3,R\right) $ algebra is
\begin{equation}
t\rightarrow \arctan \tau ~,~x\rightarrow \frac{y}{\sqrt{1+\tau ^{2}}}
\label{Ns.10}
\end{equation}%
and it easy to show that under this transformation equation (\ref{Ns.09})
becomes $\ddot{y}=0$, which is equation (\ref{NS.07}).

In order to calculate the Noether point symmetries of (\ref{NS.07}) we have
to define a Lagrangian. We recall that the Noether symmetries depend on the
particular Lagrangian we consider. Let us assume the classical Lagrangian $%
L_{1}\left( t,x,\dot{x}\right) =\frac{1}{2}\dot{x}^{2}.$ Replacing in the
Noether condition we find the associated Noether conditions
\begin{align}
\xi _{,x}& =0~,~\eta _{,xx}-\xi _{,t}=0  \label{NS.8} \\
\eta _{,t}-f_{,x}& =0~,~f_{,t}=0  \label{NS.9}
\end{align}%
whose solution gives that the Noether point symmetries of (\ref{NS.07}) for
the Lagrangian $L_{1}$ are the vector fields $X_{1},~X_{2},~X_{3},~X_{4}$
and $X_{N}=2X_{5}+X_{6}$,~with corresponding non-constant Noether functions
the $f_{3}=x~$and$~f_{4}=\frac{1}{2}x^{2}$.$~$

Furthermore, from the second theorem of Noether the corresponding first
integrals are calculated easily. The vector $X_{1}$, provides the
conservation law of energy, $X_{2}$ the conservation law of momentum, while $%
X_{3}$ gives the Galilean invariance \cite{len1}. Finally the vector fields $%
X_{4}$ and $X_{N}$ are also important because they can be used to construct
higher-order conservation laws.

A more intriguing example is the slowly lengthening pendulum whose equation
of motion in the linear approximation is\footnote{%
The time dependence in the `spring constant' is due to the length of the
pendulum's string increasing slowly \cite{werner}},%
\begin{equation}
\ddot{x}+\omega ^{2}\left( t\right) x=0.  \label{bs.03}
\end{equation}%
which also admits 8 Lie point symmetries. According to Lie's result there is
a transformation which brings (\ref{bs.03}) to the form (\ref{NS.07}). In
order to find this transformation one considers the Noether symmetries and
shows that (\ref{bs.03}) admits the quadratic first integral\cite{Lewis1}%
\begin{equation}
I=\frac{1}{2}\left\{ \left( \rho \dot{x}-\dot{\rho}x\right) ^{2}+\left(
\frac{x}{\rho }\right) ^{2}\right\} ,  \label{bs.04}
\end{equation}%
where $\rho =\rho \left( t\right) $, is a solution of the second-order
differential equation%
\begin{equation}
\ddot{\rho}+\omega ^{2}\left( t\right) \rho =\frac{1}{\rho ^{3}}.
\label{bs.05}
\end{equation}%
The first integral (\ref{bs.04}) is known as the Lewis invariant.

On the other hand, equation (\ref{bs.05}) is the well-known Ermakov-Pinney
equation \cite{Ermakov} whose solution has been given by Pinney \cite{Pinney}
and it is
\begin{equation}
\rho \left( t\right) =\sqrt{A\upsilon _{1}^{2}+2B\upsilon _{1}\upsilon
_{2}+C\upsilon _{2}^{2}}  \label{bs.06}
\end{equation}%
where $A,~B,~C$ are constants of which only two are independent, and the
functions $\upsilon _{1}\left( t\right) ,~\upsilon _{2}\left( t\right) ,$
are two linearly independent solutions of (\ref{bs.03}).

Finally, the transformation which connects the time dependent linear
equation (\ref{bs.03}) with (\ref{NS.07}) is the following
\begin{equation}
y=\frac{x}{\dot{x}}~\ ,~P=\rho \dot{x}-\dot{\rho}x~,~\tau =\int^{t}\rho
^{-2}\left( \eta \right) d\eta ,
\end{equation}%
where $\rho \left( t\right) $ is given by (\ref{bs.06}).

\subsection{The case of the $sl\left( 2,R\right) $ algebra}

The prototype system in this case is the Ermakov system defined by equations
(\ref{bs.03}) and (\ref{bs.04}) above whose Lie point symmetries span the $%
sl(2,R)$ algebra for arbitrary function $\omega \left( t\right) $, while any
solution for a specific $\omega \left( t\right) $ can be transformed to a
solution for another $\omega \left( t\right) $ by a coordinate
transformation. The Ermakov system has numerous applications in diverting
areas of Physics, see for instance \cite{rogers,schief,lee}

\bigskip Let us restrict our considerations to the autonomous case, with $%
\omega \left( t\right) =\mu ^{2}$ a constant. Equation (\ref{bs.05}) becomes%
\begin{equation}
\ddot{\rho}+\mu ^{2}\rho =\frac{1}{\rho ^{3}},
\end{equation}%
while the elements of the admitted $sl\left( 2,R\right) $ Lie algebra are%
\begin{equation*}
Z_{1}=\partial _{t}~,~Z_{2}=2t\partial _{t}+\rho \partial _{\rho }~\text{\
and~}Z_{3}=t^{2}\partial _{t}+t\rho \partial _{\rho },~\text{when }\mu =0
\end{equation*}%
and%
\begin{equation*}
Z_{1}=\partial _{t}~,~Z_{2}=\frac{1}{\mu }e^{+2i\mu t}~\partial
_{t}+e^{+2i\mu t}\rho t~\partial _{\rho }~~,~Z_{3}=\frac{1}{\mu }e^{-2i\mu
t}\partial _{t}-e^{+2i\mu t}\rho \partial _{\rho ~}~,~\text{when }\mu \neq 0.
\end{equation*}

Concerning the Noether symmetries we consider the Lagrangian
\begin{equation}
L\left( t,\rho ,\dot{\rho}\right) =\frac{1}{2}\dot{\rho}^{2}-\frac{\mu ^{2}}{%
2}\rho ^{2}-\frac{1}{2}\frac{1}{\rho ^{2}}.
\end{equation}%
and find that the Lie symmetries $Z_{1},Z_{2},Z_{3}$ satisfy the Noether
condition, hence they are also Noether point symmetries, and lead to the
quadratic first integral of energy%
\begin{equation}
\text{\thinspace }E=\frac{1}{2}\dot{\rho}^{2}+\frac{\mu ^{2}}{2}\rho ^{2}+%
\frac{1}{2}\frac{1}{\rho ^{2}}
\end{equation}%
and the time dependent first integrals
\begin{align}
I_{1}& =2tE-\rho \dot{\rho}  \label{AEP.08} \\
I_{2}& =t^{2}E-t\rho \dot{\rho}+\frac{1}{2}\rho ^{2},~\text{when }\mu =0
\label{AEP.09}
\end{align}%
or%
\begin{align}
I_{+}& =\frac{1}{2\mu }e^{+2i\mu t}E-e^{+2i\mu t}~\rho \dot{\rho}+\mu
e^{+2i\mu t}~\rho ^{2}  \label{AEP.03} \\
I_{-}& =\frac{1}{2\mu }e^{-2i\mu t}E+e^{-2i\mu t}~\rho \dot{\rho}+\mu
e^{-2i\mu t}~\rho ^{2}~,~\text{when }\mu \neq 0.  \label{AEP.04}
\end{align}

While the first integrals $I_{1},~I_{2}$ and $I_{+},~I_{-}$ are
time-dependent, we can easily construct the time-independent Lewis invariant
\cite{tsermakov}. For instance $I_{+}I_{-}$ is a time-independent first
integral.

The two dimensional system with Lagrangian%
\begin{equation}
L\left( t,\rho ,\theta ,\dot{\rho},\dot{\theta}\right) =\frac{1}{2}\dot{\rho}%
^{2}+\frac{1}{2}\rho ^{2}\dot{\theta}-\frac{\mu ^{2}}{2}\rho ^{2}-\frac{%
V\left( \theta \right) }{\rho ^{2}}  \label{aa.01}
\end{equation}%
describes the simplest generalization of the Ermakov-Pinney system in
two-dimensions. It can be shown that the Lie point symmetries $%
Z_{1},Z_{2},Z_{3}$ are Noether point symmetries of (\ref{aa.01}) with the
same first integrals. Again with the use of the time dependent Noether
integrals $I_{1},~I_{2}$ and $I_{+},~I_{-}$ we are able to construct the
autonomous conservation laws\cite{moyo}
\begin{equation}
\Phi =4I_{2}E-I_{1}^{2}=\rho ^{4}\dot{\phi}^{2}+2V\left( \phi \right)
\end{equation}%
and%
\begin{equation}
\bar{\Phi}=E^{2}-~I_{+}I_{-}=\rho ^{4}\dot{\phi}^{2}+2V\left( \phi \right)
\end{equation}%
As we shall see below the Ermakov-Pinney system and its generalizations are
used in the dark energy models \cite{tsermakov}.

\section{Symmetries of SODEs and the geometry of the underline space}

\label{sec3}

As it has been mentioned the symmetries of a SODE\ concern the kinetic
metric which is independent of the metric of the space where motion occurs.
On the other hand one of the basic principles of General Relativity (and
Newtonian Physics) is the Principle of Equivalence according to which the
trajectories of free fall are the geodesics of the space where motion
occurs. This means that the Principle of Relativity locks the Lie point
symmetries of the geodesic equations (=free fall) with the collineations of
the geometry (=metric) of space-time where the particle moves. The relation
between the Lie and Noether point symmetries and the collineations of the
space where motion occurs has been given not only for the case of geodesics
but also for a general conservative dynamical system (see \cite{p1,p2,p3,p4}%
).

Below, we briefly discuss the collineations of Riemannian manifolds and also
the results of \cite{p2} because they are used in the construction of
cosmological models.

\subsection{Collineations}

Consider a Riemannian manifold $M$ of dimension $n$ and metric $g_{ij}.$ Let
$\mathbf{A}$ be a geometric object (not necessarily a tensor) \ defined in
terms of the metric\footnote{%
This is not necessary but it is enough for the cases we consider below.}, $X$
a vector field in $F_{0}^{1}(M)$ and $\mathbf{B}$ a tensor field on $M$
which has the same number of indices as $\mathbf{A}$ and with the same
symmetries of the indices. We say that $X$ is a collineation of $\mathbf{A}$
if the following condition holds

\begin{equation}
\mathcal{L}_{X}\mathbf{A=B}  \label{L2p.2}
\end{equation}%
where $\mathcal{L}_{X}$ denotes the Lie derivative. The collineations of a
geometric object form a Lie algebra. The classification of the possible
collineations in a Riemannian space can be found in \cite{Katzin}. The most
important are the collineations given in Table \ref{Table1a}.

\begin{table}[tbp] \centering%
\caption{Collineations of the metric and of the connection in a Riemannian
space}%
\begin{tabular}{lll}
\hline\hline
\textbf{Collineation} & $\mathbf{A}$ & $\mathbf{B}$ \\ \hline
Killing vector (KV) & $g_{ij}$ & $0$ \\
Homothetic vector (HV) & $g_{ij}$ & $\psi g_{ij},\psi _{,i}=0$ \\
Conformal Killing vector (CKV) & $g_{ij}$ & $\psi g_{ij},\psi ,_{i}\neq 0$
\\
Affine collineation & $\Gamma _{jk}^{i}$ & $0$ \\
{Projective collineation (PC)} & $\Gamma _{jk}^{i}$ & $2\phi _{(,j}\delta
_{k)}^{i},$ $\phi ,_{i}\neq 0$ \\
Special Projective collineation (SPC) & $\Gamma _{jk}^{i}$ & $2\phi
_{(,j}\delta _{k)}^{i},$ $\phi ,_{i}\neq 0$ and $\phi ,_{jk}=0$ \\
\hline\hline
\end{tabular}
\label{Table1a}%
\end{table}%

Some general results concerning collineations of a Riemannian manifold $M$
of dimension $n$ are the following

a. $M$ can have at most $n(n+1)/2$ KVs and when this is the case $M$ is
called a maximally \ symmetric space. The curvature tensor of a a maximally
symmetric space is given by the expression%
\begin{equation*}
R_{abcd}=R\left( g_{ac}g_{db}-g_{ad}g_{cb}\right)
\end{equation*}%
where $R$ the curvature scalar which is a constant. Flat space is a
maximally symmetric space for which $R=0.$

b. $M$ can have at most one proper HV.

c. $M$ can have at most $\frac{\left( n+1\right) \left( n+2\right) }{2}$
proper CKVs, $n\left( n+1\right) ~$proper ACs,~and $n(n+2)$ proper PCs. A\
2-dimensional space has infinite CKVs.

d. If the metric admits a SCKV then also admits a SPC, a gradient HV and a
gradient KV \cite{HallR}

Other properties of the collineations can be found in \cite{sym5}.

What shall be important in our discussions are the collineations of the $n$-
dimensional flat space. These collineations we summarize in Table \ref%
{Table2a}. It is important to note which collineations are gradient.

\begin{table}[tbp] \centering%
\caption{The elements for the projective algebra of the Euclidian space}%
\begin{tabular}{lll}
\hline\hline
\textbf{Collineation} & \textbf{Gradient} & \textbf{Non-gradient} \\ \hline
Killing vectors (KV) & $\mathbf{S}_{I}=\delta _{I}^{i}\partial _{i}$ & $%
\mathbf{X}_{IJ}=\delta _{\lbrack I}^{j}\delta _{j]}^{i}x_{j}\partial _{i}$
\\
Homothetic vector (HV) & $\mathbf{H}=x^{i}\partial _{i}~$ &  \\
Affine Collineation (AC) & $\mathbf{A}_{II}=x_{I}\delta _{I}^{i}\partial
_{i}~$ & $\mathbf{A}_{IJ}=x_{J}\delta _{I}^{i}\partial _{i}~$ \\
Special Projective collineation (SPC) &  & $\mathbf{P}_{I}=S_{I}\mathbf{H}.~$
\\ \hline\hline
\end{tabular}
\label{Table2a}%
\end{table}%

\section{Motion and symmetries in a Riemannian space}

\label{scc1}

The equation of motion of a particle moving in a Riemannian space is given
by the SODE%
\begin{equation}
\ddot{x}^{i}+\Gamma _{jk}^{i}\dot{x}^{j}\dot{x}^{k}=F^{i}.  \label{L2P.1}
\end{equation}%
where $\Gamma _{jk}^{i}(x^{r})$ are the connection coefficients and the
field $F^{i}$ stands for the forces acting on the particle. \ The Lie
symmetries of (\ref{L2P.1}) are as follows \cite{p2}

\begin{equation}
L_{\eta }F^{i}+2\xi ,_{t}F^{i}+\xi F_{,t}^{i}+\eta ^{i},_{tt}=0
\label{PLS.09}
\end{equation}%
\begin{equation}
\left( \xi ,_{k}\delta _{j}^{i}+2\xi ,_{j}\delta _{k}^{i}\right) F^{k}+2\eta
^{i},_{t|j}-\xi ,_{tt}\delta _{j}^{i}=0  \label{PLS.10}
\end{equation}%
\begin{equation}
L_{\eta }\Gamma _{(jk)}^{i}=2\xi ,_{t(j}\delta _{k)}^{i}  \label{PLS.11a}
\end{equation}%
\begin{equation}
\xi _{(,i|j}\delta _{r)}^{k}=0.  \label{PLS.12}
\end{equation}%
From (\ref{PLS.11a}) follows that the Lie point symmetries of the SODE\ (\ref%
{L2P.1}) are generated from the special projective algebra of the space.

\subsection{Lie point symmetries of (\protect\ref{L2P.1})}

In the case where the force is autonomous and conservative, that is, $%
F^{i}=g^{ij}V_{,j}\left( x^{k}\right) $ and $V_{,j}$ is not a gradient KV of
the metric, the solution of the determining equations as also the generic
Lie symmetry vector have been given in \cite{p2} and have as follows:

\begin{itemize}
\item \emph{Case I}\textbf{\ \ }Lie point symmetries due to the affine
algebra. The resulting Lie symmetries are%
\begin{equation}
\mathbf{X}=\left( \frac{1}{2}d_{1}a_{1}t+d_{2}\right) \partial
_{t}+a_{1}Y^{i}\partial _{i}  \label{PP.02}
\end{equation}%
where $a_{1}$ and $d_{1}$ are constants, provided the potential satisfies
the condition%
\begin{equation}
\ L_{Y}V^{,i}+d_{1}V^{,i}=0.  \label{PP.03}
\end{equation}

\item \emph{Case IIa}\textbf{\ \ }The Lie point symmetries are generated by
the gradient homothetic algebra and $Y^{i}\neq V^{,i}$. The Lie point
symmetries are
\begin{equation}
\mathbf{X}=~2\psi \int T\left( t\right) dt\partial _{t}+T\left( t\right)
Y^{i}\partial _{i}  \label{PP.04}
\end{equation}%
where the function $T\left( t\right) $ is the solution of the equation $%
~T_{,tt}=a_{1}T~~$provided the potential $V(x^{i})$ satisfies the condition
\begin{equation}
\mathcal{L}_{\mathbf{Y}}V^{,i}+4\psi V^{,i}+a_{1}Y^{i}=0.  \label{PP.03a}
\end{equation}

\item \emph{Case IIb}\textbf{\ }The \ Lie point symmetries are generated by
the gradient HV $Y^{i}=\kappa V^{,i},$ where $\kappa $ is a constant. In
this case the potential is the gradient HV\ of the metric and the Lie
symmetry vectors are
\begin{equation}
\mathbf{X}=\left( -c_{1}\sqrt{\psi k}\cos \left( 2\sqrt{\frac{\psi }{k}}%
t\right) +c_{2}\sqrt{\psi k}\sin \left( 2\sqrt{\frac{\psi }{k}}t\right)
\right) \partial _{t}+\left( c_{1}\sin \left( 2\sqrt{\frac{\psi }{k}}%
t\right) +c_{2}\cos \left( 2\sqrt{\frac{\psi }{k}}t\right) \right)
H^{,i}\partial _{i}.  \label{PP.08}
\end{equation}

\item \emph{Case IIIa} \ The \ Lie point symmetries due to the proper
special projective algebra. In this case the Lie symmetry vectors are (the
index $J$ counts the gradient KVs)%
\begin{equation}
\mathbf{X}_{J}=\left( C\left( t\right) S_{J}+D\left( t\right) \right)
\partial _{t}+T\left( t\right) Y^{i}\partial _{i}  \label{PP.10}
\end{equation}%
where the functions $C(t),T(t),D(t)$ are solutions of the system of
simultaneous equations
\begin{equation}
D_{,t}=\frac{1}{2}d_{1}T~~,~~\
T_{,tt}=a_{1}T~~,~~T_{,t}=c_{2}C~~,~~D_{,tt}=d_{c}C~~,~~C_{,t}=a_{0}T
\end{equation}%
and in addition the potential satisfies the conditions
\begin{align}
\mathcal{L}_{Y}V^{,i}+2a_{0}SV^{,i}+d_{1}V^{,i}+a_{1}Y^{i}& =0  \label{PP.11}
\\
\left( S_{,k}\delta _{j}^{i}+2S,_{j}\delta _{k}^{i}\right) V^{,k}+\left(
2Y^{i}{}_{;~j}-a_{0}S\delta _{j}^{i}\right) c_{2}-d_{c}\delta _{j}^{i}& =0.
\label{PP.12}
\end{align}

\item \emph{Case IIIb}\textbf{\ \ }Lie point symmetries due to the proper
special projective algebra and $Y_{J}^{i}=\lambda S_{J}V^{,i},$ in which~$%
V^{,i}$ is a gradient HV, and $S_{J}^{,i}$ is a gradient KV. The Lie
symmetry vectors are%
\begin{equation}
X_{J}=\left( C\left( t\right) S_{J}+d_{1}\right) \partial _{t}+T(t)\lambda
S_{J}V^{,i}\partial _{i}
\end{equation}%
where the functions $C\left( t\right) $ and $T(t)$ are computed from the
relations
\begin{equation}
T_{,tt}+2C_{,t}=\lambda _{1}T~~,~~T_{,t}=\lambda _{2}C~~,~~C_{,t}=a_{0}T
\label{SP.20}
\end{equation}%
and the potential satisfies the conditions%
\begin{align}
\mathcal{L}_{Y_{J}}V^{,i}+\lambda _{1}S_{J}V^{,i}& =0  \label{SP.21} \\
C\left( \lambda _{1}S_{J}\delta _{j}^{i}+2S_{J},_{j}V^{,i}\right) +\lambda
_{2}\left( 2\lambda S_{J,j}V^{.i}+\left( 2\lambda S_{J}-a_{0}S_{J}\right)
\delta _{j}^{i}\right) & =0.  \label{SP.22}
\end{align}
\end{itemize}

The general case for time dependent conservative forces has been given in
\cite{leojmp}. In the following only the autonomous case will be considered.

\subsection{Noether point \ symmetries of (\protect\ref{L2P.1})}

In case the force is autonomous and conservative the SODE\ (\ref{L2P.1})
follows from the regular Lagrangian
\begin{equation}
L\left( t,x^{k},\dot{x}^{k}\right) =\frac{1}{2}g_{ij}\left( x^{k}\right)
\dot{x}^{i}\dot{x}^{j}-V\left( x^{k}\right) .  \label{sd.01}
\end{equation}

The Noether\ point symmetries of Lagrangian (\ref{sd.01}) as well as the
corresponding first integrals have been given in \cite{p2} and are generated
by the homothetic algebra of the metric $g_{ij}$ as follows:

\begin{itemize}
\item \emph{Case I.} The HV\ satisfies the condition:%
\begin{equation}
V_{,k}Y^{k}+2\psi _{Y}V+c_{1}=0.  \label{NSCS.14}
\end{equation}%
The Noether point symmetry vector is%
\begin{equation}
\mathbf{X}=2\psi _{Y}t\partial _{t}+Y^{i}\partial _{i},~~f=c_{1}t,
\end{equation}%
where $T\left( t\right) =a_{0}\neq 0$ and the corresponding first integral is%
\begin{equation}
\Phi =2\psi _{Y}t\emph{E}-g_{ij}Y^{i}\dot{x}^{j}+c_{1}t
\end{equation}

\item \emph{Case II.} When the metric admits the gradient KVs $S_{J}$, the
gradient HV $H^{,i}$ and the potential satisfies the condition%
\begin{equation}
V_{,k}Y^{,k}+2\psi _{Y}V=c_{2}Y+d.  \label{NSCS.17}
\end{equation}%
In this case the Noether point symmetry vector and the Noether function are%
\begin{equation}
\mathbf{X}=2\psi _{Y}\int T\left( t\right) dt\partial _{t}+T\left( t\right)
S_{J}^{,i}\partial _{i}~~~,~~f\left( t,x^{k}\right) =T_{,t}S_{J}\left(
x^{k}\right) ~+d\int Tdt.  \label{NSCS.17a}
\end{equation}%
and the functions $T(t)$ and $K\left( t\right) $ ($T_{,t}\neq 0$) are
computed from the relations%
\begin{equation}
T_{,tt}=c_{2}T~,~K_{,t}=d\int Tdt+\text{constant}  \label{NSCS.18}
\end{equation}%
where $c_{2}$ is a constant with first integral%
\begin{equation}
\bar{\Phi}=\psi _{Y}\emph{E}\int Tdt~-Tg_{ij}H^{i}\dot{x}^{j}+T_{,t}H+d\int
Tdt.
\end{equation}%
In addition to the above cases there is also the Noether point symmetry $%
\partial _{t}$ whose first integral is the energy $E$.
\end{itemize}

From the above results it is clear that in order a given dynamical system to
admit Lie / Noether point symmetries, the underlying space must admit
collineations. This means that by studying the collineations of the
underlying geometry we can infer important information for the existence or
not and also compute the Lie/Noether point symmetries. In this respect we
may say that geometry determines the evolution of the dynamical systems.

\subsection{Point symmetries of constrained Lagrangians}

The previous analysis holds for regular dynamical systems while when the
dynamical system is constrained extra conditions are introduced.

Consider the constrained Lagrangian%
\begin{equation}
\bar{L}\left( t,x^{k},\dot{x}^{k},N\right) =\frac{1}{2N}g_{ij}\dot{x}^{i}%
\dot{x}^{j}-NV\left( x^{k}\right)  \label{ssc1}
\end{equation}%
where $N=N\left( t\right) $ is a singular degree of freedom with
Euler-Lagrange equation $\frac{\partial L}{\partial N}=0$. The latter
equation is a constraint of the system. Lagrangian functions of the form of (%
\ref{ssc1}) are provided in cosmological studies.

\subsubsection{Lie point symmetries of (\protect\ref{ssc1})}

The generic Lie point symmetry vector for the Euler-Lagrange equations of (%
\ref{ssc1}) is \cite{tchr}
\begin{equation}
X_{L}=X_{N}-2a_{3}\partial _{N}  \label{ssc1.a}
\end{equation}%
where
\begin{equation}
X_{N}=\alpha _{2}\chi \left( t\right) \partial _{t}+\left( 2\alpha _{1}\tau
\left( x^{k}\right) +\alpha _{2}\chi _{,t}\left( t\right) N\right) N\partial
_{N}-\alpha _{1}\eta ^{i}\partial _{i}~,  \label{ssc1.b}
\end{equation}%
and where $\eta ^{i}$ is a CKV of the metric $g_{ij}$ with conformal factor $%
\tau \left( x^{k}\right) $ which is related with the potential by the
following condition/ constraint:%
\begin{equation}
\mathcal{L}_{\eta }V\left( x^{k}\right) +2\left( \tau \left( x^{k}\right)
+a_{3}\right) V\left( x^{k}\right) =0.  \label{ssc.3}
\end{equation}%
We note that the Lie point symmetries of the constrained Lagrangian are
generated by the elements of the conformal algebra, whereas for regular
systems these symmetries are generated by the elements of the special
projective algebra. Except that, there are also differences in the
constraint condition for the potential.

\subsubsection{Noether point symmetries of (\protect\ref{ssc1})}

In order to compute the Noether point symmetries of (\ref{ssc1}) we consider
the Noether condition (\ref{GenHolonNoether'sFinCond}). We find one Noether
point symmetry which is again the vector $X_{N}$ \cite{tchr}; the potential
satisfies the following condition / constraint:%
\begin{equation}
\mathcal{L}_{\eta }V\left( x^{k}\right) +2\tau \left( x^{k}\right) V\left(
x^{k}\right) =0.  \label{ssc.2}
\end{equation}%
The first integral defined by the Noether point symmetry $X_{N}$ is given by
the formula \cite{tchr}
\begin{equation}
\Phi ^{\ast }=\frac{1}{N}g_{ij}\eta ^{i}\dot{x}^{j}-\chi \left( t\right)
\left( \frac{\partial L}{\partial N}\right) \simeq \frac{1}{N}g_{ij}\eta ^{i}%
\dot{x}.  \label{ssc.4}
\end{equation}

The function $\Phi ^{\ast }$ is a \textquotedblleft weak\textquotedblright\
conservation law in the sense that someone has to impose the constraint
condition $\left( \frac{\partial L}{\partial N}\right) =0$, in order $\frac{%
d\Phi }{dt}=0$; this is because $\frac{d\Phi }{dt}=2\frac{\partial L}{%
\partial N}$. \ \ Furthermore we note that there is a difference in the
Noether point symmetries between the regular and the constrained Lagrangian.
For instance, while the regular Lagrangian $L\left( t,x,\dot{x}\right) =%
\frac{1}{2}\dot{x}^{2}-\frac{1}{2}x^{2}$ possesses five Noether point
symmetries \cite{notss}, for the singular Lagrangian $\bar{L}\left( t,x,\dot{%
x},N\right) =\frac{1}{2N}\dot{x}^{2}-\frac{1}{2}Nx^{2}$ we found only the $%
X_{N}$ Noether symmetry.

We note that if $a_{3}=0$ then there is only one Lie point symmetry which is
also a Noether point symmetry.

\section{Symmetries in Cosmology}

\label{sec4}

The nature of the source which drives the late-acceleration phase of the
universe is an important problem of modern cosmology. Currently the
late-acceleration phase of the universe is attributed to a perfect fluid
with negative equation of state parameter, which has been named the dark
energy. The simplest dark energy candidate is the cosmological constant
model leading to the $\Lambda $CDM cosmology. In this model the
gravitational field equations can be linearized and one is able to write the
analytic solution in closed-form. However in spite of its simplicity, the $%
\Lambda $CDM cosmological model suffers from two major problems, the fine
tuning problem and the coincidence problem \cite{Weinberg89,Peebles03,Pad03}%
. In order to overpass these problems cosmologists introduced dynamical
evolving dark energy models. In these models the dark energy fluid can be an
exotic matter source like the Chaplygin gas, quintessence, $k-$essence,
tachyons or it can be of a geometric origin provided by a modification of
Einstein's General Relativity\ \cite%
{Ratra88,Lambda4,Bas09c,Star80,bar1,bar2,bar3,bar4,mod1,mod2,mod3,mod4,mod5,mod6}%
. The dark energy components introduce new terms in the gravitational field
equations which are nonlinear or increase the degrees of freedom. The
reafter, the linearization process applied in the case of the $\Lambda $CDM
model fails and other mathematical methods must be applied in the study of
integrability of the field equations and the construction of analytic
solutions. An alternative approach is the polytropic dark matter models
which can also describe the recent acceleration of the universe through a
polytropic process \cite{kle1,kle2}

Two different groups, de Ritis et al. \cite{deRitis} and Rosquist et al.
\cite{ros1} applied independently the symmetries of differential equations
in order to construct first integrals in scalar field cosmology. In
particular, they determined the forms of the scalar field potential, which
drives the dynamics of the dark energy, in order the field equations to
admit Noether point symmetries. The classification scheme is based on an
idea proposed by Ovsiannikov \cite{Ovsi}. Since then, the classification
scheme has been applied to various dark energy models and modified theories
of gravity. Some of these classifications are complete while some others
lack mathematical completeness leading to incorrect results. \ The purpose
of the current review is to present the application of symmetries of
differential equations in modern cosmology.

A cosmological model is a relativistic model therefore requires two
assumptions: a. A specification of the metric, which is achieved mainly by
the collineations for the comoving observers we discussed above and b.
Equations of state which specify (compatible with the assumed collineations
defining the metric!) the matter of the model universe. The latter is done
by the introduction of a potential function in the action integral from
which the field equations follow. One important class of cosmological models
are the ones in which spacetime brakes in 1+3 parts, that is, the cosmic
time and the spatial universe respectively. The latter is realized
geometrically by the three dimensional spacelike hypersurfaces which are
generated by the orbits of KVs of a three dimensional Lie algebra. In 1898,
Luigi Bianchi \cite{bian} classified all possible real three dimensional
real Lie algebras in nine types. Each Lie algebra leads to a (hypersurface
orthogonal) cosmological model called a Bianchi Spatially homogeneous
cosmological model. These nine models have been studied extensively in the
literature over the years and have resulted in many important cosmological
solutions.

The principal advantage of Bianchi cosmological models is that, due to the
geometric structure of spacetime the physical variables depend only on the
time thus reducing the Einstein and the other governing equations to
ordinary differential equations \cite{rayan}. Although, the gravitational
field equations in General Relativity for the Bianchi cosmologies are
ordinary second-order differential equations, due to the existence of
nonlinear terms, exact solutions have been determined only for a few of them
\cite{bsol1,bsol2,bsol3,bsol4,bsol4a,bsol5}, while there was a debate few
years ago on the integrability or not of the Mixmaster universe (Bianchi
type IX model) \cite{bsol5,bsol6,bsol7,bsol8,bsol9}.

In order to get detailed information on these alternative models one has to
find an analytical solution of the field equations. This can be a formidable
task depending on the form of the potential function and the free parameters
that it has. The standard method to find an analytical solution is to use
Noether symmetries and compute first integrals of the field equations.
Indeed the application of symmetries of differential equations in the dark
energy models started with the use of the first Noether theorem in \cite%
{deRitis} and with the consideration of the second Noether theorem in \cite%
{ros1}. Both approaches are equivalent. Since then, Noether symmetries have
been applied to a plethora of models for the determination of first
integrals, and consequently analytical solutions. We refer the reader to
some of them \cite%
{ns00,ns01,ns02,ns03,ns04,ns05,ns06,ns07,ns08,ns09,ns10,ns11,ns12,ns13,ns14,ns15,ns16,ns17,ns18,ns19,ns20,ns21,ns22,cap03,cap04,cap05,tch10,tch20,an01,an02,an03,an04}%
. It is important to note that some of the published results are
mathematically incorrect. For instance, in \cite{mot} the authors used the
Noether conditions in order to solve the dynamical equations of the model,
and posteriori they determined the symmetries of the field equations. This
is not correct because Noether symmetries are imposed by the requirement
that they transform the Action Integral in a certain way and not as extra
conditions to the Euler-Lagrange equations.

The difficulty with the above approach is that one has to work in spacetime
where the geometry is not simple and the field equations are rather complex.
To bypass that difficulty a new scenario has been developed in which one
transforms the problem to a minisuperspace defined by the dynamical
variables through a Lagrangian which produces the field equations in that
space \cite{bas1}. Then one considers the Lagrangian in two parts. The
kinematic part which defines the kinetic metric and the remaining part which
defines the effective potential. If one knows the homothetic algebra of the
kinetic metric \cite{p2} then the application of the results of Section \ref%
{scc1} provide the Noether symmetries and the corresponding Noether first
integrals of the field equations in mini superspace. Therefore the solution
of the field equations is made possible and by the inverse transformation
one finds the solution of the original field equations in the original
dynamic variables in spacetime. This approach brought new results in various
dark energy models and modified theories of gravity \cite%
{bas2,bas3,bas4,bas5}. Some of these results are discussed in the following.

Before we enter into detailed discussion it is useful to state the action
summary of this method of work in order to provide a working tool to new
cosmologists not experience in this field.

\subsubsection{Method of work - scenario}

1. Consider the Action Integral of the model in spacetime and produce the
field equations.

2. Change variables and give a new set of field equations in the
minisuperspace of dynamical variables in a convenient form

3. Define a Lagrangian for the field equations in the mini superspace

4. Read form the Lagrangian the kinetic metric and the effective potential.
The new variables must be such that the kinetic metric will be flat or at
least one for which one knows already the homothetic algebra. This defines
the phrase "convenient form " stated in step 2 above.

5. Apply the results of section 3 to get a classification of Noether
symmetries and compute the corresponding first integrals of the field
equations in mini superspace

6. Solve these equations for the various cases of the effective potential
and other possible parameters.

7. Apply the inverse transformation and get the solution of the original
field equations in terms of the original dynamical variables in spacetime.

In the sections which follow we apply this scenario to the major
cosmological models proposed so far and give the detailed results in each
case.

\subsection{FRW spacetime and the $\Lambda $CDM cosmological model}

The FRW spacetime is a decomposable 1+3 spacetime in which the three
dimensional hypersurfaces are maximally symmetric spacelike hypersurfaces of
constant curvature which are normal to the time coordinate. The metric of a
FRW spacetime is specified modulo a function of time, the scale factor $%
a(t). $ In comoving coordinates $\{t,x,y,z\}$ it has the form%
\begin{equation}
ds^{2}=-dt^{2}+a^{2}\left( t\right) \left( dx^{2}+dy^{2}+dz^{2}\right) .
\label{let.01}
\end{equation}

In the Bianchi classification it is a Type IX spacetime. This spacetime in
classical General Relativity for comoving observers\footnote{%
For non-comovig observers can support all types of matter.} $u^{a}=\frac{%
\partial }{\partial t}$ ($u^{a}u_{a}=-1)$ can support matter which is a
perfect fluid, that is the energy momentum tensor is%
\begin{equation*}
T_{ab}=\rho u_{a}u_{b}+ph_{ab}
\end{equation*}%
where $\rho ,p$ are the energy density and the isotropic pressure of matter
as measured by the comoving observers. $h_{ab}=g_{ab}+u_{a}u_{b}$ is the
tensor projecting normal to the vector $u^{a}.~$This spacetime has been used
in the early steps of relativistic cosmology.

The first cosmological model using this spacetime was the $\Lambda $CDM
cosmology which was a vacuum spatially flat FRW spacetime with matter
generated by a non-vanishing cosmological constant $\Lambda $. For this
model Einstein field equations are%
\begin{equation}
-3a\dot{a}^{2}+2a^{3}\Lambda =0,  \label{let.03}
\end{equation}%
and
\begin{equation}
\ddot{a}+\frac{1}{2a}\dot{a}^{2}-a\Lambda =0.  \label{let.04}
\end{equation}%
whose solution is the well-known de Sitter solution $a\left( t\right)
=a_{0}e^{\sqrt{\frac{2\Lambda }{3}}t}$ which is a maximally symmetric
spacetime (not only the maximally symmetric 3d hypersurfaces). According to
earlier comments there exists a coordinate transformation which brings the
system to the linear equation (\ref{let.05}) \cite{Mahomed93}. Indeed if we
introduce the new variable $r\left( t\right) =a(t)^{3/2}$ the field equation
(\ref{let.04}) becomes \cite{hoan}%
\begin{equation}
-\frac{1}{2}\dot{r}^{2}+\frac{3}{2}\Lambda \dot{r}^{2}=0~~,~~\ddot{r}-\frac{3%
}{2}\Lambda r=0.  \label{let.05}
\end{equation}%
which is the one dimensional hyperbolic linear oscillator which admits 8 Lie
point symmetries.

Let us demonstrate the geometric scenario mentioned above in this simple
case. For this we need to find the maximum number of Noether point
symmetries admitted by the field equation (\ref{let.04}). We choose the
variables $\{t,a\}$ and have a two-dimensional mini superspace. A Lagrangian
for equation (\ref{let.04}) in the minisuperspace $\{t,a\}$ is%
\begin{equation}
L\left( t,a,\dot{a}\right) =3a\dot{a}^{2}+2\Lambda a^{3}.  \label{let.06}
\end{equation}%
from where we read the kinetic metric $s^{2}(t)=3a\dot{a}^{2}$ and the
effective potential $V(a)=-2\Lambda a^{3}.$The Noether condition (\ref%
{GenHolonNoether'sFinCond}) for the Lagrangian (\ref{let.06}) gives that it
admits five Noether point symmetries, which is the maximum number for
admitted Noether symmetries for a two-dimensional system. These Noether
symmetries are
\begin{equation*}
X_{\Lambda \left( 1\right) }=\partial _{a}~~,~~X_{\Lambda \left( 2\right) }=~%
\frac{e^{\frac{\sqrt{6\Lambda }}{2}t}}{\sqrt{a}}\partial _{a}~,~~X_{\Lambda
\left( 3\right) }=\frac{e^{-\frac{\sqrt{6\Lambda }}{2}t}}{\sqrt{a}}\partial
_{a}
\end{equation*}%
\begin{equation*}
X_{\Lambda \left( 4\right) }=e^{\frac{\sqrt{6\Lambda }}{2}t}\left( \frac{3}{%
\sqrt{\Lambda }}\partial _{t}+\sqrt{6}a\partial _{a}\right) ~,~~X_{\Lambda
\left( 5\right) }=e^{-\frac{\sqrt{6\Lambda }}{2}t}\left( \frac{3}{\sqrt{%
\Lambda }}\partial _{t}-\sqrt{6}a\partial _{a}\right) .
\end{equation*}

This simple application shows how the geometry of the kinetic metric can be
used to recognize the equivalence of well-known systems of classical
mechanics with dark energy models.

\subsection{Scalar-field cosmology}

In the case of classical General Relativity with a minimally coupled scalar
field (quintessence or phantom) the Action Integral in spacetime is
\begin{equation}
S_{M}=\int dx^{4}\sqrt{-g}\left[ R+\frac{1}{2}g_{ij}\phi ^{;i}\phi
^{;j}-V\left( \phi \right) \right] .
\end{equation}%
Assuming a spatially flat FLRW background and comoving observers the field
equations are%
\begin{equation}
-3a\dot{a}^{2}+\frac{\varepsilon }{2}a^{2}\dot{\phi}^{2}+a^{3}V\left( \phi
\right) =0  \label{SF.60e}
\end{equation}%
\begin{equation}
\ddot{a}+\frac{1}{2a}\dot{a}^{2}+\frac{\varepsilon }{4}\dot{\phi}^{2}-\frac{1%
}{2}aV=0  \label{SF.60.1}
\end{equation}%
\begin{equation}
\ddot{\phi}+\frac{3}{a}\dot{a}\dot{\phi}+\varepsilon V_{,\phi }=0.
\label{SF.60.2}
\end{equation}%
We consider the mini superspace defined by the dynamic variables $\{a,\phi
\}.$ \ A point-like Lagrangian in the mini superspace of the for field
equations (\ref{SF.60.1}) and (\ref{SF.60.2}) is
\begin{equation}
L\left( t,a,\dot{a},\phi ,\dot{\phi}\right) =-3a\dot{a}^{2}+\frac{%
\varepsilon }{2}a^{2}\dot{\phi}^{2}-a^{3}V\left( \phi \right) .
\end{equation}

To bring the Lagrangian in the "convenient form" we consider the coordinate
transformation $\{a,\phi \}$ to $\{r,\theta \}$
\begin{equation}
r=\sqrt{\frac{8}{3}}a^{3/2}\;\;,\;\;\theta =\sqrt{\frac{3\varepsilon }{8}}%
\phi \;,  \label{tran1A}
\end{equation}%
and again $\{r,\theta \}$ to $\{x,y\}$ where%
\begin{equation}
x=r\cosh \theta ~~,~y=r\sinh \theta ,
\end{equation}%
and the new variables have to satisfy the following inequality: $x\geq |y|$.
In the coordinates $\{x,y\}$ the scale factor becomes%
\begin{equation}
a=\left[ \frac{3(x^{2}-y^{2})}{8}\right] ^{1/3}.  \label{alcon}
\end{equation}

Under the coordinate transformation $\left\{ a,\phi \right\} \rightarrow
\left\{ x,y\right\} $ the point-like Lagrangian takes the simpler form
\begin{equation}
L\left( t,x,\dot{x},y,\dot{y}\right) =\frac{1}{2}\left( \dot{y}^{2}-\dot{x}%
^{2}\right) -V_{eff}(x,y)  \label{alcon1}
\end{equation}%
in which the metric in the coordinates $\{x,y\}$ is the Lorentzian 2d metric
$diag(-1,1)$ which is the metric of a flat space while the effective
potential is%
\begin{equation}
V_{eff}(x,y)=\frac{3}{8}\left( x^{2}-y^{2}\right) \tilde{V}\left( \theta
\right) .
\end{equation}

Application of the previous analysis gives the following classification of
Noether symmetries of the model for various forms of the effective potential~%
\cite{bas1}

\begin{itemize}
\item For arbitrary potential $V_{eff}\left( x,y\right) $, Lagrangian (\ref%
{alcon1}) admits the Noether point symmetry $\partial _{t}$ with first
integral the constraint equation (\ref{SF.60e}).

\item For constant potential $V\left( \theta \right) =V_{0},$ the system
admits the extra Noether symmetry $x\partial _{y}-y\partial _{x}$ with first
integral the angular momentum $r^{2}\dot{\theta}=const.$

\item For the exponential potential $V_{eff}\left( x,y\right)
=r^{2}e^{-d\theta }$, Lagrangian (\ref{alcon1}) admits an extra Noether
symmetry provided by the proper HV of the two dimensional flat space, that
is,
\begin{equation*}
X_{\left( \phi \right) 1}=2t\partial _{t}+\left( x+\frac{4}{d}y\right)
\partial _{x}+\left( y+\frac{4}{d}x\right) \partial _{y},
\end{equation*}%
with first integral
\begin{equation}
\Phi _{\left( \phi \right) 1}=\left( x+\frac{4}{d}y\right) \dot{x}-\left( y+%
\frac{4}{d}x\right) \dot{y}
\end{equation}%
\newline
while when $d=2$, the Lagrangian admits the additional symmetry $\partial
_{x}+\partial _{y}$, with corresponding Noetherian first integral
\begin{equation}
\Phi _{\left( \phi \right) 2}=\dot{x}-\dot{y}.
\end{equation}

\item Finally, when $V_{eff}(x,y)=\frac{1}{2}\left( \omega _{1}x^{2}-\omega
_{2}y^{2}\right) $, that is, $\tilde{V}(\theta )=\frac{1}{2}\left( \omega
_{1}\cosh ^{2}\left( \theta \right) -\omega _{2}\sinh ^{2}\left( \theta
\right) \right) $, the dynamical system admits four extra Noether point
symmetries
\begin{align*}
X_{\left( \phi \right) _{2}}& =\sinh \left( \sqrt{\omega _{1}}t\right)
\partial _{x}~,~X_{\left( \phi \right) _{3}}=\cosh \left( \sqrt{\omega _{1}}%
t\right) \partial _{x}~, \\
X_{\left( \phi \right) _{4}}& =\sinh \left( \sqrt{\omega _{2}}t\right)
\partial _{y}~,~X_{\left( \phi \right) _{5}}=\cosh \left( \sqrt{\omega _{2}}%
t\right) \partial _{y}~,
\end{align*}%
with corresponding first integrals%
\begin{align*}
I_{n_{2}}& =\sinh \left( \sqrt{\omega _{1}}t\right) \dot{x}-\sqrt{\omega _{1}%
}\cosh \left( \sqrt{\omega _{1}}t\right) x, \\
I_{n_{3}}& =\cosh \left( \sqrt{\omega _{1}}t\right) \dot{x}-\sqrt{\omega _{1}%
}\sinh \left( \sqrt{\omega _{1}}t\right) x, \\
I_{n_{4}}& =\sinh \left( \sqrt{\omega _{2}}t\right) \dot{y}-\sqrt{\omega _{2}%
}\cosh \left( \sqrt{\omega _{2}}t\right) y, \\
I_{n_{5}}& =\cosh \left( \sqrt{\omega _{2}}t\right) \dot{y}-\sqrt{\omega _{2}%
}\sinh \left( \sqrt{\omega _{2}}t\right) y,
\end{align*}%
The latter dynamical hyperbolic dynamical system reduces to that of the
unified dark matter potential (UDM) when $\omega _{1}=2\omega _{2}$ \cite%
{Bertaca07}. It is noticeable the amount of information one receives by the
direct application of the geometric symmetries of the kinetic metric.
\end{itemize}

To find the solution in the original dynamical variables $\left\{ a,\phi
\right\} $ we apply the inverse transformation. The result is
\begin{equation}
a^{3}\left( t\right) =\frac{3}{8}[\sinh ^{2}\left( \sqrt{\omega _{1}}%
t+\theta _{1}\right) -\frac{\omega _{1}}{\omega _{2}}\sinh ^{2}\left( \sqrt{%
\omega _{2}}t+\theta _{2}\right) ],
\end{equation}%
\begin{equation}
\phi \left( t\right) =\sqrt{\frac{8}{3\varepsilon }}\arctan h\left[ \sqrt{%
\frac{\omega _{1}}{\omega _{2}}}\frac{\sinh \left( \sqrt{\omega _{2}}%
t+\theta _{2}\right) }{\sinh \left( \sqrt{\omega _{1}}t+\theta _{1}\right) }%
\right] .
\end{equation}

\subsection{Brans-Dicke Cosmology}

The Brans-Dicke action is \cite{Brans}
\begin{equation}
S_{NM}=\int dtdx^{3}\sqrt{-g}\left[ F_{0}\psi ^{2}R-\frac{1}{2}\bar{g}%
_{ij}\psi ^{;i}\psi ^{;j}+\bar{V}\left( \psi \right) \right]
\end{equation}%
where $F_{0}$ is related to the Brans-Dicke parameter.

In the case of the spatially flat FRW background and comoving observers, the
point-like Lagrangian in the mini superspace defined by the variables $%
\{a,\psi \}$ which describes the gravitational field equations is%
\begin{equation}
L\left( t,a,\dot{a},\psi ,\dot{\psi}\right) =6F_{0}\psi ^{2}a\dot{a}%
^{2}-12F_{0}\psi a^{2}\dot{a}\dot{\psi}-\frac{1}{2}a^{3}\dot{\psi}%
^{2}+a^{3}V\left( \psi \right) .  \label{CLN.21}
\end{equation}

If one performs the coordinate transformation $\{a,\psi \}$ to $\{r,\theta
\} $ by the equations%
\begin{equation}
a\simeq r^{\frac{2}{3}}~\ ,~~\theta \simeq \ln \psi ,
\end{equation}
Lagrangian (\ref{CLN.21}) becomes
\begin{equation}
L\left( t,r,\dot{r},\theta ,\dot{\theta}\right) =e^{k\theta }\left(
-dr^{2}+r^{2}d\theta ^{2}\right) -r^{2}V\left( \theta \right)  \label{CLN.30}
\end{equation}%
from which we have that the kinetic metric of the minisuperspace is the
conformally flat Lorentzian 2d metric $e^{k\theta }\left(
-dr^{2}+r^{2}d\theta ^{2}\right) $ whose symmetry algebra depends on the
values $\left\vert k\right\vert \neq 1$ $\left\vert k\right\vert =1$ while
the effective potential is $V_{effec.}=-r^{2}V\left( \theta \right) .$

We consider cases.

\subsubsection{Case $\left\vert k\right\vert \neq1$. \newline
}

For $\left\vert k\right\vert \neq 1$ the homothetic algebra of the
minisuperspace consists of the gradient KVs

\begin{align}
K^{1}& =\frac{e^{\left( 1-k\right) \theta }r^{k}}{N_{0}^{2}}\left( -\partial
_{r}+\frac{1}{r}\partial _{\theta }\right)  \label{KV.1} \\
K^{2}& =\frac{e^{-\left( 1+k\right) \theta }r^{-k}}{N_{0}^{2}}\left(
\partial _{r}+\frac{1}{r}\partial _{\theta }\right)  \label{KV.2}
\end{align}%
\qquad the non gradient KV
\begin{equation}
K^{3}=r\partial _{r}-\frac{1}{k}\partial _{\theta }  \label{KV.3}
\end{equation}%
and the gradient HV%
\begin{equation}
H^{i}=\frac{1}{N_{0}^{2}\left( k^{2}-1\right) }\left( -r\partial
_{r}+k\partial _{\theta }\right) ~,~H\left( r,\theta \right) =\frac{1}{2}%
\frac{r^{2}e^{2k\theta }}{k^{2}-1}.  \label{KV.4}
\end{equation}

The symmetry classification provides the following results \cite{bas4}

\begin{itemize}
\item For arbitrary potential $V\left( \theta \right) $, the dynamical
system admits the Noether point symmetry $\partial _{t}$.

\item For $V\left( \theta \right) =V_{0}e^{2\theta }$ there are two
additional Noether symmetries $K^{1},~tK^{1}$ with first integrals%
\begin{equation}
I_{1}=\frac{d}{dt}\left( \frac{r^{1+k}e^{\left( 1+k\right) \theta }}{\left(
k+1\right) }\right) ~,~I_{2}=t\frac{d}{dt}\left( \frac{r^{1+k}e^{\left(
1+k\right) \theta }}{\left( k+1\right) }\right) -\left( \frac{%
r^{1+k}e^{\left( 1+k\right) \theta }}{\left( k+1\right) }\right) .
\label{CLN.75aa}
\end{equation}

\item For $V\left( \theta \right) =V_{0}e^{2\theta }-\frac{mN_{0}^{2}}{%
2\left( k^{2}-1\right) }e^{2k\theta }$ there are two additional Noether
symmetries $e^{\pm \sqrt{m}t}K^{1}$, with corresponding first integrals%
\begin{equation}
I_{\pm }^{\prime }=e^{\pm \sqrt{m}t}\left[ \frac{d}{dt}\left( \frac{%
r^{1+k}e^{\left( 1+k\right) \theta }}{\left( k+1\right) }\right) \mp \sqrt{m}%
\left( \frac{r^{1+k}e^{\left( 1+k\right) \theta }}{\left( k+1\right) }%
\right) \right] .  \label{CLN.99}
\end{equation}

\item For $V\left( \theta \right) =V_{0}e^{-2\theta }$,we have the extra
Noether symmetries $K^{2},~tK^{2}$ with Noether first integrals%
\begin{equation}
J_{1}=\frac{d}{dt}\left( \frac{r^{1-k}e^{-\left( 1-k\right) \theta }}{k-1}%
\right) ~,~J_{2}=t\frac{d}{dt}\left( \frac{r^{1-k}e^{-\left( 1-k\right)
\theta }}{k-1}\right) -\frac{r^{1-k}e^{-\left( 1-k\right) \theta }}{k-1}.
\label{CLN.100}
\end{equation}

\item For $V\left( \theta \right) =V_{0}e^{-2\theta }-\frac{mN_{0}^{2}}{%
2\left( k^{2}-1\right) }e^{2k\theta }$, we have the extra Noether symmetries
$e^{\pm \sqrt{m}t}K^{2}$ $m=$constant, with Noether first integrals%
\begin{equation}
J_{\pm }^{^{\prime }}=e^{\pm \sqrt{m}t}\left( \frac{d}{dt}\left( \frac{%
r^{1-k}e^{-\left( 1-k\right) \theta }}{k-1}\right) \mp \sqrt{m}\frac{%
r^{1-k}e^{-\left( 1-k\right) \theta }}{k-1}\right) .
\end{equation}

\item For the potential $V\left( \theta \right) =V_{0}e^{2k\theta }$ the
additional symmetry is the vector field $K^{3}$ with first integrals as
given by the second theorem of Noether
\begin{equation}
I_{3}=\frac{re^{2k\theta }}{k}\left( k\dot{r}+r\dot{\theta}\right) .
\end{equation}

\item For $V\left( \theta \right) =V_{0}e^{-2\frac{\left( k^{2}-2\right) }{k}%
\theta },k^{2}-2\neq 0$ the extra Noether symmetries are $2t\partial
_{t}+H^{i}~,~t^{2}\partial _{t}+tH^{i}$ with first integrals%
\begin{equation}
I_{H_{1}}=-\frac{d}{dt}\left( \frac{1}{2}\frac{r^{2}e^{2k\theta }}{k^{2}-1}%
\right) ~,~I_{H_{2}}=-t\frac{d}{dt}\left( \frac{1}{2}\frac{r^{2}e^{2k\theta }%
}{k^{2}-1}\right) +\frac{1}{2}\frac{r^{2}e^{2k\theta }}{k^{2}-1}.
\label{CLN.100c}
\end{equation}%
We note that in this case the system is the Ermakov-Pinney dynamical system
(because it admits the Noether symmetry algebra the $sl(2,R),\ $hence the
Lie symmetry algebra is at least $sl(2,R))$ .

\item For $V\left( \theta \right) =V_{0}e^{-2\frac{\left( k^{2}-2\right) }{k}%
\theta }-\frac{N_{0}^{2}m}{k^{2}-1}e^{2k\theta }~$ , $k^{2}-2\neq 0$ we have
the Noether symmetries $\frac{2}{\sqrt{m}}e^{\pm \sqrt{m}t}\partial _{t}\pm
e^{\pm \sqrt{m}t}H^{i}$ , $m=$constant with Noether first integrals%
\begin{equation}
I_{+,-}=e^{\pm 2\sqrt{m}t}\left( \mp \frac{d}{dt}\left( \frac{1}{2}\frac{%
r^{2}e^{2k\theta }}{k^{2}-1}\right) +2\sqrt{m}\left( \frac{1}{2}\frac{%
r^{2}e^{2k\theta }}{k^{2}-1}\right) \right) .  \label{CLN.105}
\end{equation}%
For this potential the Noether symmetries form the $sl\left( 2,R\right) $
Lie algebra, i.e. the dynamical system is the two dimensional Kepler-Ermakov
system.

\item The case $V\left( \theta \right) =0$ corresponds the two dimensional
free particle in flat space and the dynamical system admits seven additional
Noether symmetries.
\end{itemize}

\subsubsection{Case $\left\vert k\right\vert =1$}

We have to consider two cases i.e. $k=1$ and $k=-1.$ It is enough to
consider the case $k=1$, because the results for $k=-1$ are obtained
directly from those for $k=1$ if we make the substitution $\theta _{\left(
k=-1\right) }=-\bar{\theta}.$

For $k=1,~$the homothetic algebra of the minisuperspace$~$is given by the
vector fields $K_{k=1}^{1,2}$ of (\ref{KV.1},\ref{KV.2}) and the vector
field
\begin{equation}
K_{k=1}^{3}=-r\left( \ln \left( re^{-\theta }\right) -1\right) \partial
_{r}+\ln \left( re^{-\theta }\right) \partial _{\theta }.  \label{CLN.68}
\end{equation}%
Hence, the symmetry classification provides the following cases

\begin{itemize}
\item For arbitrary potential $V\left( \theta \right) $, the dynamical
system admits the Noether symmetry $\partial _{t}$.

All the rest cases admit additional symmetries.

\item If $V\left( \theta \right) =V_{0}e^{2\theta }~$ we have the extra
Noether symmetries $K^{1}~,~tK^{1}$ with Noether first integrals the (\ref%
{CLN.75aa}) with $k=1$.

\item If $V\left( \theta \right) =V_{0}e^{2\theta }-\frac{m}{2}\theta
e^{2\theta }$ we have the Noether symmetries $e^{\pm \sqrt{m}t}K^{1}$ with
Noether first integrals the (\ref{CLN.99}) with $k=1$.

\item Noether symmetries generated by the KV $K^{2}$.

\item If $V\left( \theta \right) =V_{0}e^{-2\theta }$ then we have the
Noether symmetries $K^{2}~,~tK^{2}$ with Noether first integrals%
\begin{equation*}
I_{2}^{\prime }=\frac{d}{dt}\left( \theta -\ln r\right) ~,~I_{2}^{\prime }=t%
\left[ \frac{d}{dt}\left( \theta -\ln r\right) \right] -\left( \theta -\ln
r\right) .
\end{equation*}

\item If $V\left( \theta \right) =V_{0}e^{-2\theta }-\frac{1}{4}pe^{2\theta
}\,$ then we have the Noether symmetries $K^{2}~,~tK^{2}$ with Noether first
integrals%
\begin{equation*}
I_{1}^{\prime }=\frac{d}{dt}\left( \theta -\ln r\right) -pt~,~I_{2}^{\prime
}=t\left[ \frac{d}{dt}\left( \theta -\ln r\right) \right] -\left( \theta
-\ln r\right) -\frac{1}{2}pt^{2}.
\end{equation*}

\item If $V\left( \theta \right) =0$ then the system becomes the free
particle and admits seven extra Noether symmetries.
\end{itemize}

The exact solutions of the models and their physical properties can be found
in \cite{bas4}. The results from the classification analysis are presented
in Tables \ref{bdpoint1} and \ref{bdpoint2}. For the notation of the
admitted Lie algebra we follow the {Mubarakzyanov Classification Scheme }%
\cite{Mubarakzyanov63a,Mubarakzyanov63b,Mubarakzyanov63c}

\begin{table}[tbp] \centering%
\caption{Noether symmetry classification for the Brans-Dicke action in a
spatially flat FLRW spacetime (I)}%
\begin{tabular}{ccccc}
\hline\hline
$\left\vert \mathbf{k}\right\vert $ & \textbf{Potential} & \textbf{\#
Symmetries} & \textbf{Lie Algebra} & \textbf{Symmetries} \\ \hline
& $V\left( \theta\right) $ & $1$ & $A_{1}$ & $\partial_{t}~,~$ \\
$\neq1$ & $V_{0}e^{2\theta}$ & $3$ & $A_{1}\otimes_{s}\left( 2A_{1}\right) $
& $\partial_{t},~K^{1},~tK^{1}$ \\
$\neq1$ & $V_{0}e^{2\theta}-\frac{mN_{0}^{2}}{2\left( k^{2}-1\right) }%
e^{2k\theta}$ & $3$ & $A_{1}\otimes_{s}\left( 2A_{1}\right) $ & $%
\partial_{t},~e^{\pm\sqrt{m}t}K^{1}$ \\
$\neq1$ & $V_{0}e^{-2\theta}$ & $3$ & $A_{1}\otimes_{s}\left( 2A_{1}\right) $
& $\partial_{t},~K^{2},~tK^{2}$ \\
$\neq1$ & $V_{0}e^{-2\theta}-\frac{mN_{0}^{2}}{2\left( k^{2}-1\right) }%
e^{2k\theta}$ & $3$ & $A_{1}\otimes_{s}\left( 2A_{1}\right) $ & $%
\partial_{t},~e^{\pm\sqrt{m}t}K^{21}$ \\
$\neq1$ & $V_{0}e^{2k\theta}$ & $2$ & $2A_{1}$ & $\partial_{t},~K^{3}$ \\
$\neq1$ & $V_{0}e^{-2\frac{\left( k^{2}-2\right) }{k}\theta},k^{2}-2$ & $3$
& $Sl\left( 2,R\right) $ & $\partial_{t},~2t\partial_{t}+H^{i}~,~t^{2}%
\partial_{t}+tH^{i}$ \\
$\neq1$ & $V_{0}e^{-2\frac{\left( k^{2}-2\right) }{k}\theta}-\frac{N_{0}^{2}m%
}{k^{2}-1}e^{2k\theta}~$ , $k^{2}-2\neq0$ & $3$ & $Sl\left( 2,R\right) $ & $%
\partial_{t},~\frac{2}{\sqrt{m}}e^{\pm\sqrt{m}t}\partial_{t}\pm e^{\pm\sqrt{m%
}t}H^{i}$ \\ \hline\hline
\end{tabular}
\label{bdpoint1}%
\end{table}%

\begin{table}[tbp] \centering%
\caption{Noether symmetry classification for the Brans-Dicke action in a
spatially flat FLRW spacetime (II)}%
\begin{tabular}{ccccc}
\hline\hline
$\left\vert \mathbf{k}\right\vert $ & \textbf{Potential} & \textbf{\#
Symmetries} & \textbf{Lie Algebra} & \textbf{Symmetries} \\ \hline
$=1$ & $V_{0}e^{2\theta}$ & $3$ & $A_{1}\otimes_{s}\left( 2A_{1}\right) $ & $%
\partial_{t},~K^{1},~tK^{1}$ \\
$=1$ & $V_{0}e^{2\theta}-\frac{m}{2}\theta e^{2\theta}$ & $3$ & $%
A_{1}\otimes_{s}\left( 2A_{1}\right) $ & $\partial_{t},~e^{\pm\sqrt{m}%
t}K^{1} $ \\
$=1$ & $V_{0}e^{-2\theta}$ & $3$ & $A_{1}\otimes_{s}\left( 2A_{1}\right) $ &
$\partial_{t},~K^{2},~tK^{2}$ \\
$=1$ & $V_{0}e^{-2\theta}-\frac{1}{4}pe^{2\theta}\,$ & $3$ & $A_{1}\otimes
_{s}\left( 2A_{1}\right) $ & $\partial_{t},~K^{2},~tK^{2}$ \\ \hline\hline
\end{tabular}
\label{bdpoint2}%
\end{table}%

\subsection{$f\left( R\right) $-gravity}

$f\left( R\right) $-gravity (in the metric formalism) is a fourth-order
theory where the Action Integral in spacetime is
\begin{equation}
S=\int d^{4}x\sqrt{-g}f\left( R\right)  \label{action1}
\end{equation}%
In the case of FRW background and comoving observers the resulting field
equations follow from the Lagrangian\cite{Sot10}
\begin{equation}
L\left( t,a,\dot{a},R,\dot{R}\right) =6af^{^{\prime }}~\dot{a}%
^{2}+6a^{2}f^{^{\prime \prime }}~\dot{a}\dot{R}+a^{3}\left( f^{^{\prime
}}R-f\right) -6Kaf^{\prime }  \label{lanfr}
\end{equation}%
where $K=0,\pm 1$ is the spatial curvature of the FRW spacetime, and a prime
denotes derivative with respect to the dynamical parameter $R$, that is $%
f^{\prime }\left( R\right) =\frac{df\left( R\right) }{dR}$. Since $f\left(
R\right) $ theory can be written as a special case of Brans-Dicke theory,
the so-called O'Hanlon gravity \cite{Sot10}, the results will be common with
that of the previous analysis. However for completeness we present them
below.

The classification scheme provides the following cases \cite{bas2}:

\begin{itemize}
\item For arbitrary function $f\left( R\right) $, there exists the
autonomous symmetry $\partial _{t}$ , which derives the constraint equation.

\item For $f\left( R\right) =R^{\frac{3}{2}}$, the theory admits the
additional Noether symmetries%
\begin{equation}
K_{1}=2t\partial _{t}+\frac{4}{3}~\partial _{a}-\frac{9}{2}\frac{f^{\prime }%
}{f^{\prime \prime }}\partial _{R},
\end{equation}%
\begin{equation}
K_{2}=\frac{1}{a}\partial _{a}-\frac{1}{a^{2}}\frac{f^{\prime }}{f^{\prime
\prime }}\partial _{R}~,~K_{2}^{\ast }=t\left( \frac{1}{a}\partial _{a}-%
\frac{1}{a^{2}}\frac{f^{\prime }}{f^{\prime \prime }}\partial _{R}\right) ,
\end{equation}%
with first integrals%
\begin{equation}
\Phi _{1}=6a^{2}\dot{a}\sqrt{R}+6\frac{a^{3}}{\sqrt{R}}\dot{R},
\end{equation}%
\begin{equation}
\Phi _{2}=\frac{d}{dt}\left( a\sqrt{R}\right) ~,~\Phi _{2}^{\ast }=t\frac{d}{%
dt}\left( a\sqrt{R}\right) -a\sqrt{R}.
\end{equation}

\item For $f\left( R\right) =R^{\frac{7}{8}}$ and $K=0$, the theory admits
the additional Noether symmetries%
\begin{equation}
K_{3}=2t\partial _{t}+\frac{a}{2}~\partial _{a}+\frac{1}{2}\frac{f^{\prime }%
}{f^{\prime \prime }}\partial _{R}~,~K_{3}^{\ast }=t^{2}\partial
_{t}+t\left( \frac{a}{2}~\partial _{a}+\frac{1}{2}\frac{f^{\prime }}{%
f^{\prime \prime }}\partial _{R}\right) ,
\end{equation}%
with corresponding first integrals
\begin{equation}
\Phi _{3}=\frac{d}{dt}\left( a^{3}R^{-\frac{1}{8}}\right) ~~,~\Phi
_{3}^{\ast }=t\frac{d}{dt}\left( a^{3}R^{-\frac{1}{8}}\right) -a^{3}R^{-%
\frac{1}{8}}.
\end{equation}

\item The power-law theory $f\left( R\right) =R^{n}$ (with $n\neq 0,1,\frac{3%
}{2},\frac{7}{8}$) $\ $and for $K=0,$ or with $K$ arbitrary and $n=2$, the
system admits the extra symmetry%
\begin{equation}
K_{1}^{\ast }=2t\partial _{t}+\left( \frac{4n}{3}-\frac{2}{3}\right)
a\partial _{t}-3\frac{f^{\prime }}{f^{\prime \prime }}\partial _{R},
\end{equation}%
with first integral%
\begin{equation}
\Phi _{1}^{\ast }=a^{2}R^{n-1}\dot{a}\left( 2-n\right) +\frac{1}{2}%
a^{3}R^{n-2}\dot{R}\left( 2n-1\right) \left( n-1\right) .
\end{equation}

\item For $f\left( R\right) =(R-2\Lambda )^{3/2}$ the extra Noether
symmetries are%
\begin{equation}
K_{\left( \pm \right) 2}=e^{\pm \sqrt{m}t}\left( \frac{1}{a}\partial _{a}-%
\frac{1}{a^{2}}\frac{f^{\prime }}{f^{\prime \prime }}\partial _{R}\right) ,
\end{equation}%
with first integrals%
\begin{equation}
\Phi _{\left( \pm \right) 2}=e^{\pm \sqrt{m}t}\left( \frac{d}{dt}\left( a%
\sqrt{R-2L}\right) \mp 9\sqrt{m}a\sqrt{R-2\Lambda }\right) .  \label{NI.b2}
\end{equation}

\item Finally, when $f\left( R\right) =(R-2\Lambda )^{7/8}$ the field
equations admit the Noether symmetries%
\begin{equation}
K_{\left( \pm \right) 4}=\pm \frac{1}{\sqrt{m}}e^{\pm 2\sqrt{m}t}\partial
_{t}+e^{\pm 2\sqrt{m}t}\left( \frac{a}{2}~\partial _{a}+\frac{1}{2}\frac{%
f^{\prime }}{f^{\prime \prime }}\partial _{R}\right) ,
\end{equation}%
with corresponding first integrals
\begin{equation}
\Phi _{\left( \pm \right) 4}=\frac{d}{dt}\left( a^{3}\left( R-2\Lambda
\right) ^{-\frac{1}{8}}\right) \mp \frac{1}{2}\sqrt{m}a^{3}\left( R-2\Lambda
\right) ^{-\frac{1}{8}}.
\end{equation}
\end{itemize}

In Table \ref{frclas1}, we collect the results of the classification scheme
for $f\left( R\right) $-gravity.

\begin{table}[tbp] \centering%
\caption{Noether symmetry classification for $f(R)$ in FLRW spacetime}%
\begin{tabular}{ccccc}
\hline\hline
\textbf{Sp. Curv. }$K$ & $\mathbf{f}\left( \mathbf{R}\right) $ & \textbf{\#
Symmetries} & \textbf{Lie Algebra} & \textbf{Symmetries} \\ \hline
$=0,\pm1$ & Arbitrary & $1$ & $A_{1}$ & $\partial_{t}~,~$ \\
$=0,\pm1$ & $R^{\frac{3}{2}}$ & $4$ & $\left( 2A_{1}\right) \otimes
_{s}\left( 2A_{1}\right) $ & $\partial_{t},~K_{1},~K_{2},~K_{2}^{\ast}$ \\
$=0$ & $R^{\frac{7}{8}}$ & $3$ & $Sl\left( 2,R\right) $ & $\partial
_{t},~K_{3},~K_{3}^{\ast}$ \\
$=0$ & $R^{n}$ (with $n\neq0,1,\frac{3}{2},\frac{7}{8}$) & $2$ & $2A_{1}$ & $%
\partial_{t},~K_{1}^{\ast}$ \\
$=1$ & $R^{2}$ & $2$ & $A_{1}\otimes_{s}\left( 2A_{1}\right) $ & $%
\partial_{t},~K_{1\left( n=2\right) }^{\ast}$ \\
$=0,\pm1$ & $(R-2\Lambda)^{3/2}$ & $3$ & $A_{1}\otimes_{s}\left(
2A_{1}\right) $ & $\partial_{t},~K_{\left( \pm\right) 2}$ \\
$=0$ & $(R-2\Lambda)^{7/8}$ & $3$ & $Sl\left( 2,R\right) $ & $\partial
_{t},~K_{\left( \pm\right) 4}$ \\ \hline\hline
\end{tabular}
\label{frclas1}%
\end{table}%

\subsection{Two-scalar field cosmology}

We consider now a two-scalar field cosmological model in General Relativity
with Action Integral%
\begin{equation}
S=\int dx^{4}\sqrt{-g}\left( R-\frac{1}{2}g_{ij}\left( \Phi ^{C}\right) \Phi
^{A,i}\Phi ^{B,i}+V\left( \Phi ^{C}\right) \right) ,  \label{TF.01}
\end{equation}%
where $H_{AB}$ describes the coupling between the two scalar fields $\Phi
^{A}=\left( \phi ,\psi \right) $ in the kinematic part. Moreover, we assume
the metric tensor $H_{AB}$ to be a maximally symmetric metric of constant
curvature \cite{bas5}. In such a scenario it is not possible to define new
fields in order to remove the coupling in the kinematic part.

Assuming again a spatially flat FRW spacetime and comoving observers the
field equations are%
\begin{eqnarray}
-3a\dot{a}^{2}+\frac{1}{2}a^{3}H_{AB}\dot{\Phi}^{A}\dot{\Phi}%
^{B}+a^{3}V\left( \Phi ^{C}\right) &=&0, \\
\ddot{a}+\frac{1}{2a}\dot{a}^{2}+\frac{a}{4}H_{AB}\dot{\Phi}^{A}\dot{\Phi}%
^{B}-\frac{1}{2}aV &=&0, \\
\ddot{\Phi}^{A}+\frac{3}{2a}\dot{a}\dot{\Phi}^{A}+\tilde{\Gamma}_{BC}^{A}%
\dot{\Phi}^{B}\dot{\Phi}^{C}+H^{AB}V_{,B} &=&0,
\end{eqnarray}%
where $\tilde{\Gamma}_{BC}^{A}$ are the connection coefficients for the
metric $H_{AB}\left( \Phi ^{C}\right) .$

In the mini superspace defined by the variables $\{a,\Phi \}$ we introduce
the new variables $\{u,\phi \}$ by the requirements $a=\left( \frac{3}{8}%
\right) ^{\frac{1}{3}}u^{\frac{2}{3}},$ $diag\left( 1,e^{2\phi }\right)
=h_{AB}\dot{\Phi}^{A}\dot{\Phi}^{B}$ and the field equations become%
\begin{align}
-\frac{1}{2}\dot{u}^{2}+\frac{1}{2}u^{2}\left( \dot{\phi}^{2}+e^{2\phi }~%
\dot{\psi}^{2}\right) +u^{2}V\left( \phi ,\psi \right) & =0, \\
\ddot{u}+u\dot{\phi}^{2}+ue^{2\phi }~\dot{\psi}^{2}-2uV& =0, \\
\ddot{\phi}+\frac{2}{u}\dot{u}\dot{\phi}-e^{2\phi }\dot{\psi}^{2}+V_{,\phi
}& =0, \\
\ddot{\psi}+\frac{2}{u}\dot{u}\dot{\psi}+2\dot{\phi}\dot{\psi}+e^{-2\phi
}V_{,\psi }& =0.
\end{align}

These follow form the point-like Lagrangian%
\begin{equation}
L\left( t,u,\dot{u},\phi ,\dot{\phi},\psi ,\dot{\psi}\right) =-\frac{1}{2}%
\dot{u}^{2}+\frac{1}{2}u^{2}\left( \dot{\phi}^{2}+e^{2\phi }~\dot{\psi}%
^{2}\right) -u^{2}V\left( \phi ,\psi \right)
\end{equation}

which defines the 3d flat Lorentzian kinetic metric $-\frac{1}{2}\dot{u}+%
\frac{1}{2}u^{2}\left( \dot{\phi}^{2}+e^{2\phi }~\dot{\psi}^{2}\right) $ and
the effective potential $V_{eff.}=u^{2}V\left( \phi ,\psi \right) .$

The kinetic metric admits a seven dimensional homothetic algebra consisting
of the three gradient KVs (translations)%
\begin{equation*}
K^{1}=-\frac{1}{2}\left( e^{\phi }\left( 1+\psi ^{2}\right) +e^{-\phi
}\right) \partial _{u}+\frac{1}{2u}\left( e^{\phi }\left( 1+\psi ^{2}\right)
-e^{-\phi }\right) \partial _{\phi }+\frac{1}{u}\psi e^{-\phi }\partial
_{\psi },
\end{equation*}%
\begin{align*}
K^{2}& =-\frac{1}{2}\left( e^{\phi }\left( 1-\psi ^{2}\right) -e^{-\phi
}\right) \partial _{u}+\frac{1}{2u}\left( e^{\phi }\left( 1-\psi ^{2}\right)
+e^{-\phi }\right) \partial _{\phi }-\frac{1}{u}\psi e^{-\phi }\partial
_{\psi }, \\
K^{3}& =-\psi e^{\theta }\partial _{u}+\frac{1}{u}\psi e^{\phi }\partial
_{\phi }+\frac{1}{u}e^{-\phi }\partial _{\psi }
\end{align*}%
with corresponding gradient functions $S_{\left( 1-3\right) }$ given by%
\begin{align*}
S_{\left( 1\right) }& =\frac{1}{2}u\left( e^{\phi }\left( 1+\psi ^{2}\right)
+e^{-\phi }\right) ,~ \\
S_{\left( 2\right) }& =\frac{1}{2}u\left( e^{\phi }\left( 1-\psi ^{2}\right)
-e^{-\phi }\right) , \\
S_{\left( 3\right) }& =u\psi e^{\theta },
\end{align*}%
the three non-gradient KVs (rotations) which span the $SO\left( 3\right) $
algebra
\begin{equation*}
X_{12}=\partial _{\psi }~,~\ X_{23}=\partial _{\phi }+\psi \partial _{\psi
},X_{13}=\psi \partial _{\phi }+\frac{1}{2}\left( \psi ^{2}-e^{2\phi
}\right) \partial _{\phi }
\end{equation*}%
and the gradient proper HV
\begin{equation*}
H_{V}=u\partial _{u}~,~~\psi _{H_{V}}=1.
\end{equation*}

The classification of the Noether symmetries for the various potentials $%
V\left( \phi ,\psi \right) $ is as follows \cite{bas5}:

\begin{itemize}
\item For arbitrary potential $V\left( \phi ,\psi \right) $, the field
equations admit the Noether symmetry $\partial _{t}$ which provides the
constraint equation of General Relativity.

\item For $V\left( \phi ,\psi \right) =0$, the dynamical system is maximally
symmetric and admits in total twelve Noether symmetries.

\item For $V\left( \phi ,\psi \right) =V\left( \phi \right) $, there exists
the additional Noether symmetry, the vector field $X_{12}$, with
conservation law the angular momentum on the two dimensional sphere, that is
\begin{equation*}
\Phi _{12}=e^{2\phi }\dot{\psi}.
\end{equation*}

\item For $V_{A}\left( \phi ,\psi \right) =\frac{\omega _{0}^{2}}{2}u^{2}+%
\frac{\mu ^{2}}{2\left( 1-a_{0}^{2}\right) }\left( S_{\left( \mu \right)
}+a_{0}S_{\left( \nu \right) }\right) ^{2}-\frac{\omega _{3}^{2}}{2}%
S_{\left( \sigma \right) }^{2}~,~a_{0}\neq 1$, the system admits six
additional Noether symmetries given by the vector fields%
\begin{equation*}
T_{1}\left( t\right) K^{1}~,~T_{2}\left( t\right) K^{2}~,T_{3}\left(
t\right) K^{3}
\end{equation*}%
where
\begin{equation*}
T_{,tt}^{A}=\omega _{~\delta }^{\gamma }T^{\delta }~,~\omega _{~\delta
}^{\gamma }=diag\left( \left( \omega _{1}\right) ^{2},\left( \omega
_{2}\right) ^{2},\left( \omega _{3}\right) ^{3}\right)
\end{equation*}%
and $\mu ,\nu ,\sigma =1,2,3$. $\ $The corresponding Noether integrals are
expressed as follows%
\begin{equation}
I_{C}^{\gamma }=T_{\gamma }\frac{d}{dt}S_{\left( \gamma \right) }-T_{\gamma
,t}S_{\left( \gamma \right) }.
\end{equation}

However, when two constants $\omega _{A}$ are equal, for instance, $\omega
_{\mu }=\omega _{\nu }$, then the dynamical system admits an extra Noether
symmetry. That is, it admits the rotation normal to the plane defined by the
axes $x^{\mu },x^{\nu }$ given by the vector
\begin{equation*}
X=x^{\nu }\partial _{\mu }-\varepsilon x^{\mu }\partial _{\nu }
\end{equation*}%
where $\varepsilon =-1$ if $x^{\nu }/x^{\mu }=x$ and $\varepsilon =1$ if $%
x^{\nu }/x^{\mu }\neq 1$. $\ $

\item For the potential is $V_{B}\left( \phi ,\psi \right) =\frac{\omega
_{0}^{2}}{2}u^{2}+\frac{\mu ^{2}}{2\left( 1-a_{0}^{2}\right) }\left(
S_{\left( \mu \right) }+a_{0}S_{\left( \nu \right) }\right) ^{2}-\frac{%
\omega _{3}^{2}}{2}S_{\left( \sigma \right) }^{2}~,~a_{0}\neq 1$, the
dynamical system admits the extra Noether symmetries%
\begin{equation*}
\bar{T}\left( t\right) \left( K^{\mu }+a_{0}K^{\nu }\right) ~~,~T^{\prime
}\left( t\right) K^{\sigma }~,~T^{\ast }\left( t\right) \left( a_{0}K^{\mu
}+K^{\nu }\right)
\end{equation*}%
where the functions $T,~T^{\prime }$ and $\bar{T}$ are given by the linear
second-order differential equations
\begin{equation}
\bar{T}_{,tt}=\left( \mu ^{2}+\omega _{0}^{2}\right) \bar{T}~,~T_{\sigma
,tt}=\left( \omega _{3}^{2}+\omega _{0}^{2}\right) T_{\sigma
}~,~T_{,tt}^{\ast }=\omega _{0}^{2}\bar{T},
\end{equation}%
and $\mu ,\nu ,\sigma =1,2,3$. $\ $Finally, the corresponding Noether
integrals are expressed as follows%
\begin{equation}
\Phi _{1a2}=\bar{T}\frac{d}{dt}\left( S_{\left( \mu \right) }+a_{0}S_{\left(
\nu \right) }\right) -\bar{T}_{,t}\left( S_{\left( \mu \right)
}+a_{0}S_{\left( \nu \right) }\right) ,
\end{equation}%
\begin{equation}
\Phi _{3}=T_{\sigma }\frac{d}{dt}S_{\left( \sigma \right) }-T_{\sigma
,t}S_{\left( \sigma \right) },
\end{equation}%
\begin{equation}
\Phi _{a12}=T^{\ast }\frac{d}{dt}\left( a_{0}S_{\left( \mu \right)
}+S_{\left( \nu \right) }\right) -T_{,t}^{\ast }\left( a_{0}S_{\left( \mu
\right) }+S_{\left( \nu \right) }\right) .
\end{equation}
\end{itemize}

In both last cases, from the admitted algebras of Lie symmetries it is easy
to recognize that the gravitational field equations can be linearized.
Indeed for the potential $V_{A}\left( \phi ,\psi \right) $ under the
coordinate transformation%
\begin{align}
x& =\frac{1}{2}u\left( e^{\phi }\left( 1+\psi ^{2}\right) +e^{-\phi }\right)
,  \label{tr.01} \\
~~y& =\frac{1}{2}u\left( e^{\phi }\left( 1-\psi ^{2}\right) -e^{-\phi
}\right) ,  \label{tr.02} \\
z& =u\psi e^{\phi },  \label{tr.03}
\end{align}%
the field equations become%
\begin{align}
\ddot{x}-\left( \omega _{1}\right) ^{2}x& =0, \\
\ddot{y}-\left( \omega _{2}\right) ^{2}y& =0, \\
\ddot{z}-\left( \omega _{3}\right) ^{2}z& =0,
\end{align}%
\begin{equation}
-\frac{1}{2}\dot{x}^{2}+\frac{1}{2}\dot{y}+\frac{1}{2}\dot{z}^{2}+\frac{%
\omega _{1}^{2}}{2}x^{2}-\frac{\omega _{2}^{2}}{2}y^{2}-\frac{\omega _{3}^{2}%
}{2}z^{2}=0.
\end{equation}%
which is the three-dimensional \textquotedblleft
unharmonic-oscillator\textquotedblright .

On the other hand, for the potential $V_{B}\left( \phi ,\psi \right) $, we
perform the additional transformation
\begin{equation}
x=\left( w+v\right) ~,~~y=\frac{1}{a_{0}}\left( w-v\right) ~,~z=z,
\label{tr.11}
\end{equation}%
and the field equations are linearized as follows%
\begin{equation}
\ddot{w}-\left( \mu ^{2}+\omega _{0}^{2}\right) w=0,  \label{CB.8}
\end{equation}%
\begin{equation}
\ddot{v}+\frac{a_{0}^{2}+1}{a_{0}^{2}-1}\mu ^{2}w-\omega _{0}^{2}\nu =0,
\label{CB.9}
\end{equation}%
\begin{equation}
\ddot{z}-\left( \omega _{3}^{2}+\omega _{0}^{2}\right) z=0,  \label{CB.10}
\end{equation}%
\begin{align}
0& =\frac{1}{2}\left[ \left( \frac{1}{a_{0}^{2}}-1\right) \dot{w}^{2}-\left(
\frac{1}{a_{0}^{2}}+1\right) dwdv+\left( \frac{1}{a_{0}^{2}}-1\right) dv^{2}+%
\frac{1}{2}z^{2}\right] +  \notag \\
& \ -\frac{2\mu ^{2}}{\left( a_{0}^{2}-1\right) }w^{2}-\frac{1}{2}\left(
\omega _{3}^{2}+\omega _{0}^{2}\right) z^{2}+\frac{\omega _{0}^{2}}{2}\left(
\left( w+v\right) ^{2}-\frac{1}{a_{0}^{2}}\left( w-v\right) ^{2}\right) .
\label{CB.11}
\end{align}

\subsection{Galilean cosmology}

The cubic Galilean cosmological model in a spatially flat FRW spacetime with
comoving observers is defined by the Lagrangian \cite{gal02}
\begin{equation}
L\left( a,\dot{a},\phi ,\dot{\phi}\right) =3\,a\,\dot{a}^{2}-\frac{1}{2}\,{a}%
^{3}\dot{\phi}^{2}+{a}^{3}V(\phi )+g(\phi )a^{2}\dot{a}\,\dot{\phi}^{3}-%
\frac{g^{\prime }(\phi )}{6}{a}^{3}\,\dot{\phi}^{4}.  \label{lag}
\end{equation}

From the symmetry condition we should determine two functions, $V\left( \phi
\right) $ and $g\left( \phi \right) .$ Indeed, we find that when \cite{ns19}%
\begin{equation}
V(\phi )=V_{0}e^{-\lambda \phi }\quad \text{and}\quad g(\phi
)=g_{0}e^{\lambda \phi }  \label{pot}
\end{equation}%
Lagrangian (\ref{lag}) admits the Noether point symmetries%
\begin{equation}
X_{1}=\partial _{t}~,~X_{2}=t\partial _{t}+\frac{a}{3}\partial _{a}+\frac{2}{%
\lambda }\partial _{\phi },  \label{ns}
\end{equation}%
which form the $2A_{1}$ Lie algebra.

Noether symmetry $X_{1}$ provides as first integral the constraint equation,
while $X_{2}$ gives the first integral%
\begin{equation}
\Phi _{2}=-\left( {2\,a^{2}\dot{a}{-}\frac{{2}}{\lambda }a^{3}\dot{\phi}}%
\right) {+g_{0}e^{\lambda \phi }a^{3}\dot{\phi}^{3}-}\frac{6}{\lambda }%
g_{0}a^{2}e^{\lambda \phi }\dot{a}\dot{\phi}^{2}.  \label{con01}
\end{equation}

Furthermore, the same first integral exists in the limit in which $V_{0}=0$.
In addition, we remark that when the universe is dominated by the potential
of the scalar field then \thinspace $g\left( \phi \right) \rightarrow 0$,
and the model reduces to that of a minimally coupled scalar field.

\section{Higher-Order Symmetries in Cosmology}

\label{sec6}

In the previous section we presented classification of cosmological models
based on point transformations. However, these are not the only cases where
first integrals are used. Indeed one is possible to extend the
classification scheme by applying non-point symmetries such as the contact
symmetries.

In particular for Lagrangians of the form (\ref{sd.01}) it has been found
that the vector field $X=K_{j}^{i}\left( t,q^{k}\right) \dot{x}^{i}\partial
_{i}$ is a contact symmetry for the Action Integral iff the following
conditions are satisfied \cite{kalotas}%
\begin{equation}
K_{\left( ij;k\right) }=0,  \label{LB.07}
\end{equation}%
\begin{equation}
K_{ij,t}=0~~,~f_{,t}=0,  \label{LB.08}
\end{equation}%
\begin{equation}
K^{ij}V_{j}+f_{,i}=0.  \label{LB.09}
\end{equation}%
The latter conditions follow directly from the application of the weak
Noether condition. From the symmetry condition (\ref{LB.08}) it follows that
$K_{j}^{i}=K_{j}^{i}\left( q^{k}\right) $ and $f=f\left( q^{k}\right) $. \
Furthermore, the second-theorem of Noether provides the first integral%
\begin{equation}
I=K_{ij}\dot{x}^{i}\dot{x}^{j}-f\left( x^{i}\right) .
\end{equation}

Condition (\ref{LB.07}) means that the second rank tensor $K_{j}^{i}\left(
q^{k}\right) $ is a Killing tensor of order 2 of the metric $g_{ij}$.
Condition (\ref{LB.09}) is a constraint relating the potential with the
Killing tensor $K^{ij}$ and the Noether function $f$. Application of contact
symmetries in cosmological studies can be found in \cite%
{ros1,contact1,contact2,contact3,contact4}.

{From the following results we shall see that the classification
scheme according to the admitted contact symmetries, include the dynamical
systems which admit point symmetries, as also, provide new integrable
dynamical systems. More specifically, as we let more freedom in the admitted
symmetry vectors then we can find new integrable dynamical systems. }

{As far as concerns the contact symmetries, their application in
classical mechanic is important and they can explain the Runge-Lenz vector
for the Kepler problem, the Lewis invariant for the Ermakov-Pinney system
and many others. As far as concerns the application in cosmology from
previous studies \cite{contact1,contact2,contact3,contact4} new integrable
models and new analytic solutions for the evolution of the universe have
been determined. }

In the following we present the results for the classification of contact
symmetries in scalar-field cosmology and $f\left( R\right) $-gravity.

\subsection{Scalar-field cosmology from contact symmetries}

In the polar coordinates (\ref{tran1A}) the Lagrangian of the field
equations in scalar field cosmology becomes%
\begin{equation}
L\left( r,\theta ,\dot{r},\dot{\theta}\right) =-\frac{1}{2}\dot{r}^{2}+\frac{%
1}{2}r^{2}\dot{\theta}^{2}-r^{2}V\left( \theta \right)  \label{CS.08}
\end{equation}%
while the Killing tensors of rank two for two-dimensional flat space in
Cartesian coordinates $\left\{ x,y\right\} $ are
\begin{equation}
K_{ij}=%
\begin{pmatrix}
c_{1}y^{2}+2c_{2}y+c_{3} & c_{6}-c_{1}yx-c_{2}x-c_{4}y \\
c_{6}-c_{1}yx-c_{2}x-c_{4}y & c_{1}x^{2}+2c_{4}x+c_{5}%
\end{pmatrix}%
.
\end{equation}

Thus condition (\ref{LB.09}) provides the following cases \cite{contact1}:

Case A: For the hyperbolic Potential $\ V\left( \theta \right)
=c_{1}+c_{2}\cosh ^{2}\theta $ the field equations admit the contact
symmetry
\begin{equation}
X_{1}=-\left( \cosh ^{2}\theta \dot{r}+\frac{1}{2}r\sinh \left( 2\theta
\right) \dot{\theta}\right) \partial _{r}+\frac{1}{r}\left( \frac{1}{2}\sinh
\left( 2\theta \right) \dot{r}+r\sinh \theta ~\dot{\theta}\right) \partial _{%
\dot{r}}
\end{equation}%
with corresponding Noether Integral%
\begin{equation}
I_{1}=\left( \cosh \theta \dot{r}+r\sinh \theta ~\dot{\theta}\right)
^{2}-2r^{2}\left( c_{1}+c_{2}\right) \cosh ^{2}\theta .
\end{equation}%
In the special case where \thinspace $c_{2}=3c_{1}$ the field equations
admit the second contact symmetry
\begin{equation}
\bar{X}=-r^{2}\sinh \theta ~\dot{\theta}\partial _{r}+\left( \sinh \theta
\dot{r}+2r\cosh \theta ~\dot{\theta}\right) \partial _{\theta }
\label{Pot2l}
\end{equation}%
with corresponding Noether Integral%
\begin{equation}
\bar{I}_{1}=\left( r^{2}\sinh \theta ~\dot{r}\dot{\theta}+r^{3}\cosh \theta ~%
\dot{\theta}^{2}\right) +2c_{1}r^{3}\cosh \theta \sinh ^{2}\theta
\end{equation}

Case B: For the potential $V\left( \theta \right) =c_{1}\left( 1+3\cosh
^{2}\theta \right) +c_{2}\left( 3\cosh \theta +\cosh ^{3}\theta \right) $
there exists the contact symmetry%
\begin{equation}
I_{2}=\left( r^{2}\sinh \theta ~\dot{r}\dot{\theta}+r^{3}\cosh \theta ~\dot{%
\theta}^{2}\right) +r^{3}\sinh ^{2}\theta \left( 2c_{1}\cosh \theta
+c_{2}\left( 1+\cosh ^{2}\theta \right) \right)
\end{equation}%
with Noether Integral%
\begin{equation}
I_{2}=\left( r^{2}\sinh \theta ~\dot{r}\dot{\theta}+r^{3}\cosh \theta ~\dot{%
\theta}^{2}\right) +r^{3}\sinh ^{2}\theta \left( 2c_{1}\cosh \theta
+c_{2}\left( 1+\cosh ^{2}\theta \right) \right)
\end{equation}

Case C: For potential $V\left( \theta \right) =c_{1}\left( 1-3\sinh
^{2}\theta \right) +c_{2}\left( 3\sinh \theta -\sinh ^{3}\theta \right) \,\ $%
\ the admitted contact symmetry is
\begin{equation}
\bar{X}_{2}=-r^{2}\cosh \theta ~\dot{\theta}\partial _{r}+\left( \cosh
\theta ~\dot{r}+2r\sinh \theta ~\dot{\theta}\right) \partial _{\theta }
\end{equation}%
with corresponding Noether Integral
\begin{equation}
\bar{I}_{2}=\left[ \cosh \theta ~\dot{r}+r\sinh \theta ~\dot{\theta}\right)
r^{2}\dot{\theta}-r^{3}\cosh ^{2}\theta \left( 2c_{1}\sinh \theta
-c_{2}\left( 1-\sinh ^{2}\theta \right) \right] .
\end{equation}%
This potential is equivalent to case B under the transformation $\theta =%
\bar{\theta}+i\frac{\pi }{2}.$

Case D: For $V\left( \theta \right) =c_{1}+c_{2}e^{2\theta }$ the admitted
contact symmetry is
\begin{equation}
X_{3}=-e^{2\theta }\left( \dot{r}+r\dot{\theta}\right) \partial _{r}+\frac{%
e^{2\theta }}{r}\left( \dot{r}+r\dot{\theta}\right) \partial _{\theta }
\end{equation}%
with corresponding Noether Integral
\begin{equation}
I_{3}=e^{2\theta }\left( \left( \dot{r}+r\dot{\theta}\right)
^{2}-2r^{2}c_{1}\right) .
\end{equation}%
Moreover, when $c_{1}=0$ the dynamical system admits the additional contact
symmetry%
\begin{equation}
X_{3}=-r^{2}e^{\theta }\dot{\theta}\partial _{r}+e^{\theta }\left( \dot{r}+2r%
\dot{\theta}\right) \partial _{\theta }  \label{Pot4l}
\end{equation}%
with corresponding Noether Integral%
\begin{equation}
\bar{I}_{3}=r^{2}e^{\theta }\left( \dot{r}\dot{\theta}+r\dot{\theta}%
^{2}\right) +\frac{2}{3}c_{2}r^{3}e^{3\theta }.
\end{equation}

Case E: Finally, for the potential
\begin{equation}
V\left( \theta \right) =c_{1}e^{2\theta }+c_{2}e^{3\theta }  \label{Pot.4}
\end{equation}%
the field equations admit the first integral%
\begin{equation}
I_{3}=r^{2}e^{\theta }\left( \dot{r}\dot{\theta}+r\dot{\theta}^{2}\right)
+r^{3}e^{3\theta }\left( \frac{2}{3}c_{1}+c_{2}e^{\theta }\right)
\label{Pot.4II}
\end{equation}%
generated by the contact symmetry \ref{Pot4l}. It is important to note that
in all cases the results remain the same under the transformation $\theta
\rightarrow -\theta $.

\subsection{$f\left( R\right) $-gravity from contact symmetries}

\label{fr111}

Without loss of generality we define $\phi =f^{\prime }\left( R\right) $,
where now $f\left( R\right) $-gravity can be written in its equivalent from
as a Brans-Dicke scalar field cosmological model. Specifically the
Lagrangian of the field equations is written equivalently as
\begin{equation}
L\left( a,\dot{a},\phi ,\phi \right) =6a\phi \dot{a}^{2}+6a^{2}\dot{a}\dot{%
\phi}+a^{3}V\left( \phi \right) ,  \label{fr.13}
\end{equation}%
where%
\begin{equation}
V\left( \phi \right) =\left( f^{\prime }R-f\right) \text{ or }V\left(
f^{\prime }\left( R\right) \right) =\left( f^{\prime }R-f\right) .
\label{fr.14}
\end{equation}

The classification in terms of the contact symmetries provides the following
five cases for the potential $V\left( \phi \right) $ and the corresponding
first integrals.

Case A: For $V_{I}\left( \phi \right) =V_{1}\phi +V_{2}\phi ^{3}$, the field
equations admit the quadratic first integral%
\begin{equation}
I_{I}=3\left( \phi \dot{a}+a\dot{\phi}\right) ^{2}-V_{1}~a^{2}\phi ^{2}
\label{fr.15}
\end{equation}%
generated by the KT $K_{22}^{ij}$.

Case B: For $V_{II}\left( \phi \right) =V_{1}\phi -V_{2}\phi ^{-7},~$ the
field equations admit the quadratic first integral%
\begin{equation}
I_{II}=3a^{4}\left( \phi \dot{a}-a\dot{\phi}\right) ^{2}+4V_{2}a^{6}\phi
^{-6},  \label{fr.16}
\end{equation}

Case C:\ For $V_{III}\left( \phi \right) =V_{1}-V_{2}\phi ^{-\frac{1}{2}},~$%
the field equations admit the quadratic first integral
\begin{equation}
I_{III}=6a^{3}\dot{a}\left( a\dot{\phi}-\phi \dot{a}\right) -a^{5}\left(
\frac{3}{5}V_{1}-V_{2}\phi ^{-\frac{1}{2}}\right) .  \label{fr.17}
\end{equation}

Case D: For $V_{IV}\left( \phi \right) =V_{1}\phi ^{3}+V_{2}\phi ^{4},$ the
field equations admit the quadratic first integral
\begin{equation}
I_{IV}=12a^{2}\left( a^{2}\dot{\phi}^{2}-\phi ^{2}\dot{a}^{2}\right) +\left(
a\phi \right) ^{4}\left( 3V_{1}+4V_{2}\phi \right) .  \label{fr.18}
\end{equation}

Case E: For $V_{V}\left( \phi \right) =V_{1}\left( \phi ^{3}+\beta \phi
\right) +V_{2}\left( \phi ^{4}+6\beta \phi ^{2}+\beta ^{2}\right) ,~$the
field equations admit the quadratic first integral
\begin{align}
I_{V}& =12a^{2}\left[ \left( \beta -\phi ^{2}\right) \dot{a}^{2}+a^{2}\dot{%
\phi}^{2}\right] +  \notag \\
& -a^{4}\left( \beta -\phi ^{2}\right) \left[ V_{1}\left( \beta +3\phi
^{2}\right) +4V_{2}\left( 3\beta \phi +\phi ^{3}\right) \right] .
\label{fr.19}
\end{align}

However in order to derive the function $f\left( R\right) $ one has to solve
the Clairaut equation (\ref{fr.14}). For the above cases, Clairaut equation
has a closed-form solution only for some particular forms of $V\left( \phi
\right) $. The analytic forms of $f\left( R\right) $ functions which admit
contact symmetries are presented in Table \ref{frcc}.

\begin{table}[tbp] \centering%
\caption{Analytic forms of $f(R)$ theory where the field equations admits
contact symmetries}%
\begin{tabular}{cc}
\hline\hline
\textbf{Potential }$V\left( \phi \right) $ & Function $f\left( R\right) $ \\
\hline
$V_{I}\left( \phi \right) $ & $\left( R-V_{1}\right) ^{\frac{3}{2}}$ \\
$V_{II}\left( \phi \right) $ & $\left( R-V_{1}\right) ^{\frac{7}{8}},$ \\
$V_{III}\left( \phi \right) $ & $R^{\frac{1}{3}}-V_{1}$ \\
$V_{IV}\left( \phi \right) ~,~V_{1}=0$ & $R^{4}$ \\
$V_{IV}\left( \phi \right) ~,~V_{2}=0$ & $R^{\frac{3}{2}}$ \\
$V_{V}\left( \phi \right) ,~V_{1}=\pm 4V_{2}\sqrt{\beta }$ & $\mp \sqrt{%
\beta }R+R^{\frac{4}{3}}$ \\ \hline\hline
\end{tabular}%
\label{frcc}%
\end{table}%

\section{Conclusions}

In this work we discussed the dark energy problem using the classification
of the cosmological models which are based on the FRW background for
comoving observers using the tool of Noether symmetries. We discussed the
definitions for the Lie and Noether symmetries (point and generalized) for
conservative holonomic dynamical systems. Moreover, we established the
relation of Lie and Noether symmetries with the properties of the underlying
geometry for singular and regular dynamical systems. In particular for
regular dynamical systems we found that the generators of Noether symmetries
are the elements of the homothetic algebra of the mini superspace defined by
the dynamical variables of the system whereas for the singular systems the
generators of Noether symmetries are constructed by the CKVs of the mini
superspace.

These geometric results have been used to develop a geometric scenario based
on the admitted Noether symmetries of the mini superspace metric which leads
to the classification scheme of the dark energy models. We demonstrated the
application of this scenario to the most well known cosmological models
including the modified theories of gravity and derived in many of them
analytical cosmological solutions. This scenario is not limited to the case
of cosmological models and can be applied to other areas of study of
dynamical equations especially in the case of general holonomic systems.

{However, a direct question is, how the integrable cosmological
models determined by the application of symmetries describe the real
universe? The integrable cosmological models are not sensitive on the
initial conditions and they can be used as toy models for the study of
various phases in the evolution of the universe. For instance, different
potentials in scalar-field cosmology, or different functions in }$f\left(
R\right) ${-gravity provides different cosmological evolution which
can describe various eras in the evolution of the universe.}

{For instance, the }$f_{V}\left( R\right) ${\ model which
determined in Section (\ref{fr111}), for some specific initial conditions
and when }$V_{1}>4V_{2}\sqrt{\beta }${, it has been found that the
Hubble function is given by \cite{contact2} }%
\begin{equation}
\left( \frac{H\left( a\right) }{H_{0}}\right) ^{2}\simeq \Omega
_{r0}a^{-4}+\Omega _{m0}a^{-3}+\Omega _{\Lambda }  \label{fr.57}
\end{equation}%
{which corresponds to a universe filed with radiation, dust fluid and
cosmological constant. In a similar way it was found that the majority of
the integrable models can describe various eras of special interests \cite%
{ns16,bas1,bas2,contact1,contact2,bar01A}.}

{Various cosmological constraints for the integrable models which
discussed before have been performed in \cite{ns16,an03,contact1,contact2}
and references therein. In particular, from the data analysis have been
found that the integrable models describe well the late-time acceleration
phase of the universe. }

\bigskip%

\subsubsection*{Acknowledgement}%

AP acknowledges financial support of FONDECYT grant no. 3160121


\end{document}